\documentclass[a4paper,12pt]{article}
\pdfoutput=1

\usepackage{amsmath, amssymb, authblk, a4wide, bm, cancel, color, graphicx, multirow, subfig, enumerate, appendix, youngtab, cite, hyperref, bbold, verbatim, slashed}

\newcommand{\tr}[1]{\mathop{\rm tr}\left[#1\right]}

\newcommand{\nbrack}[1]{\left(#1\right)}
\newcommand{\sbrack}[1]{\left[#1\right]}

\newcommand{\norm}[1]{\left|#1\right|}
\newcommand{\pfrac}[2]{\left(\frac{#1}{#2}\right)}

\def\fund{\tiny\Yvcentermath1\yng(1)}

\def\asym{\tiny\Yvcentermath1\yng(1,1)}
\def\asymT{\tiny\Yvcentermath1\yng(1,1,1)}

\def\be{\begin{equation}}
\def\ee{\end{equation}}
\def\ba{\begin{eqnarray}}
\def\ea{\end{eqnarray}}
\def\cA{{\cal A}}
\def\cJ{{\cal J}}
\def\cL{{\cal L}}

\def\cO{{\cal O}}

\def\p0{{\phantom{+}0}}

\def\uno{\mbox{1 \kern-.59em {\rm l}}}

\numberwithin{equation}{section}

\begin{document}

\title{
\vspace{-3cm}
\rightline{\sc\small UMN-TH-3401/14}
\vspace{2cm}
\Large{\textbf{The Unnatural Composite Higgs}}}
\author[1]{\small{\bf James Barnard}\thanks{\texttt{james.barnard@unimelb.edu.au}}}
\author[2]{\small{\bf Tony Gherghetta}\thanks{\texttt{tgher@umn.edu}}}
\author[3]{\small{\bf Tirtha Sankar Ray}\thanks{\texttt{tirthasankar.ray@gmail.com}}}
\author[1]{\small{\bf Andrew Spray}\thanks{\texttt{andrew.spray@coepp.org.au}}}
\affil[1]{ARC Centre of Excellence for Particle Physics at the Terascale, School of Physics, 
The University of Melbourne, Victoria 3010, Australia }
\affil[2]{School of Physics and Astronomy, University of Minnesota, Minneapolis, MN 55455, USA}
\affil[3]{Department of Physics, Indian Institute of Technology Kharagpur, 721 302, India}
\date{}
\maketitle

\vspace{-1cm}
\begin{abstract}
\baselineskip=15pt
\noindent
Composite Higgs models can trivially satisfy precision-electroweak and flavour constraints by simply having a large spontaneous symmetry breaking scale, $f\gtrsim10$~TeV\@.  This produces a `split' spectrum, where the strong sector resonances have masses greater than 10 TeV and are separated from the pseudo Nambu-Goldstone bosons, which remain near the electroweak scale.  Even though a tuning of order $10^{-4}$ is required to obtain the observed Higgs boson mass, the big hierarchy problem remains mostly solved.  Intriguingly, models with a fully-composite right-handed top quark also exhibit improved gauge coupling unification.  By restricting ourselves to models which preserve these features we find that the symmetry breaking scale cannot be arbitrarily raised, leading to an upper bound $f\lesssim100$--1000~TeV\@.  This implies that the resonances may be accessible at future colliders, or indirectly via rare-decay experiments.  Dark matter is identified with a pseudo Nambu-Goldstone boson, and we show that the smallest coset space containing a stable, scalar singlet and an unbroken $SU(5)$ symmetry is $SU(7)/SU(6)\times U(1)$.  The colour-triplet pseudo Nambu-Goldstone boson also contained in this coset space is metastable due to a residual symmetry.  It can decay via a displaced vertex when produced at colliders, leading to a distinctive signal of unnaturalness.
\end{abstract}
\newpage
\tableofcontents
\section{Introduction}

A natural Higgs sector generically requires new states at the electroweak scale.  The fact that these states remain conspicuously absent at the LHC suggests that our notion of naturalness is perhaps misguided, and that there is some other reason for the Higgs to be so light.  Indeed, the Standard Model (SM) is valid all the way up to the Planck scale and the new states demanded by naturalness arguments are not mandatory for consistency of the theory. That said, the SM is clearly not a complete description of nature.  New physics is certainly required to explain dark matter and the baryon asymmetry of the universe, and is strongly implied by phenomena such as gauge coupling unification and the fermion mass hierarchy.  Many models attempting to explain the big hierarchy, between the electroweak and Planck scales, provide new physics to explain some of these features. This suggests that they still play a crucial role, even though there remains a little hierarchy between the electroweak and new physics scales, that is yet to be understood.

Composite Higgs models provide an elegant solution to the big hierarchy problem as the Higgs sector is replaced by a new, strongly coupled sector, so is rendered insensitive to any physics above the compositeness scale.  In this strong sector a global symmetry group is spontaneously broken at a scale $f$ and the Higgs is identified as one of the associated Nambu-Goldstone bosons~\cite{Kaplan:1983fs,Kaplan:1983sm,Dugan:1984hq}.  In modern incarnations the global symmetry is explicitly broken by linear couplings of elementary, SM states to operators in the strong sector, thereby inducing electroweak symmetry breaking and generating a mass for the Higgs~\cite{Contino:2003ve, Agashe:2004rs}.  To obtain the correct electroweak vacuum expectation value (VEV), $v$, a tuning of order $v^2/f^2$ is required in terms contributing to the Higgs potential.

In addition to the Higgs, the strong dynamics produces heavier resonances near the scale $f$.  These contribute to precision-electroweak and flavour observables and inevitably force an uneasy compromise between minimising the electroweak VEV tuning, which prefers $f\sim v$, and agreement with experimental results, which requires $f\gg v$.  To relieve this tension a custodial symmetry can be introduced to cancel the dominant contributions to the electroweak $T$ parameter~\cite{Agashe:2003zs}, while flavour-changing processes can be mitigated by additional symmetries~\cite{Rattazzi:2000hs,Cacciapaglia:2007fw, Santiago:2008vq}.  Even then, the absence of any resonances at the LHC means that composite Higgs models are becoming increasingly tuned.

An alternative approach is to simply increase the scale $f$ such that all constraints from precision-electroweak and flavour observables are trivially satisfied.  This can be achieved by simply requiring that $f\gtrsim 10$~TeV\@.  With such a large value there is no need for a custodial symmetry or any special flavour structure in the strong sector.  The resonance masses are now quite heavy, greater than around 10~TeV, whereas the pseudo Nambu-Goldstone bosons (pNGBs) remain near the electroweak scale.  The spectrum is therefore `split' in a similar way to the spectrum of split supersymmetry~\cite{Wells:2003tf,ArkaniHamed:2004fb, Arvanitaki:2012ps, ArkaniHamed:2012gw}.  Of course, there is now a sizeable tuning in the Higgs sector, of order $10^{-4}$, due to some cancellation in the unknown strong dynamics, but the big hierarchy problem remains mostly solved.

While it would seem that the scale $f$ could be pushed to arbitrarily high values, leaving just a light Higgs at low energy without any other observable consequences, this is not necessarily the case.  It is well known that composite Higgs models naturally explain the fermion mass hierarchy through the idea of partial compositeness~\cite{Kaplan:1991dc}.  A fully-composite right-handed top quark leads to an order-one top quark Yukawa coupling, whereas the remaining SM fermions are mostly elementary and have hierarchically smaller Yukawa couplings~\cite{Gherghetta:2000qt}.  Remarkably, exactly the same principle helps to unify the SM gauge couplings when the model is extended to a grand unified theory (GUT), as the running of the gauge couplings is modified by new states associated with the fully-composite right-handed top quark~\cite{Agashe:2005vg}.  Arbitrarily raising the scale $f$, and therefore the masses of these new states, worsens the accuracy of the gauge coupling unification.  Requiring acceptable unification of the SM gauge couplings puts an upper bound on the symmetry breaking scale, $f\lesssim 100$--1000~TeV, with a range in the upper bound following from the uncertainty in estimating the higher-loop contributions.  This implies that resonance masses cannot be arbitrarily heavy and may be accessible at a 100~TeV collider, or lead to rare decay processes.

If the resonances are heavier than 10~TeV it is difficult for any of them to provide a realistic dark matter candidate~\cite{Agashe:2004ci, Agashe:2004bm}.  However, enlarging the coset space beyond that of the minimal composite Higgs model leads to an alternative possibility; one of the extra pNGBs can play the role of dark matter~\cite{Frigerio:2012uc}.  If such a model is to be embedded into a GUT the unbroken global symmetry group in the strong sector must at least contain $SU(5)$, so that all one-loop corrections to the gauge couplings from the strong sector are universal.  The smallest coset space allowing for an unbroken $SU(5)$ symmetry and a stable, SM-singlet scalar is then $SU(7)/SU(6)\times U(1)$.  This contains twelve Nambu-Goldstone bosons, a complex $\bf5$ of $SU(5)$ providing the usual Higgs GUT multiplet and a complex singlet providing the dark matter candidate, and is the symmetry breaking pattern we will focus on in this paper.  An interesting feature of this coset space is that gauge interactions generate the leading order contribution to the quartic term in the Higgs potential so, at leading order, the Higgs mass ends up being proportional to the $W$-boson mass. 

In models based on the $SU(7)/SU(6)\times U(1)$ coset space the SM fermions couple to fermionic, strong sector operators forming complete $SU(6)$ multiplets.  In the particular model we will consider, the right-handed top quark is a composite state that lives in the $\bf15$ of $SU(6)$.  The remaining twelve states have not been observed so must each be paired with a conjugate, elementary fermion partner such that the Dirac pair obtains a mass of order $f$.  These exotic states, which we will refer to as top companions, $\chi$, decay promptly and may be searched for at a 100~TeV collider.

On the other hand there is also a colour-triplet pNGB\@.  This state is present in any composite, $SU(5)$ GUT as the pNGBs necessarily come in complete GUT multiplets; in this case a $\bf5$ of $SU(5)$ (the usual Higgs GUT multiplet).  In our model this scalar triplet has suppressed decays because of a residual $Z_2$ symmetry.  Hence it must decay to two singlets via a dimension-six operator, often leading to a decay with a displaced vertex.  This striking experimental signal is similar to that of displaced gluino decays in models with split supersymmetry.

In summary we see that, by allowing for a large scale of spontaneous symmetry breaking, $f\gtrsim 10$~TeV, a composite Higgs model can be constructed that evades all current precision-electroweak and flavour constraints.  The model is necessarily tuned in order to account for the little hierarchy, $f\gg v$.  This ``unnatural" composite Higgs model can then be characterised by the following features.
\begin{itemize}
\item \textbf{Minimal coset space:} The minimal coset space incorporating gauge coupling unification and a scalar singlet dark matter candidate is $SU(7)/SU(6)\times U(1)$.  The Nambu-Goldstone bosons form a complex singlet, $S$, and a complex ${\bf 5}=(T,D)$ of $SU(5)$ with the following features:
\begin{itemize}
\item \textbf{Higgs:} The doublet, $D$, is the Higgs boson whose mass is proportional to the $W$-boson mass.
\item \textbf{Dark matter:} The singlet, $S$, provides a scalar dark matter candidate whose stability is guaranteed by a $U(1)$ symmetry, arising from enlarging the global symmetry group from $SU(7)$ to $U(7)$.
\item \textbf{Triplet decay:} The colour-triplet, $T$, is metastable due to a residual symmetry.  This often leads to a displaced vertex when it is produced at colliders.
\end{itemize}
\item \textbf{Fermion mass hierarchy:} The right-handed top quark is fully composite and the remainder of the SM fermions are mostly elementary.  Exotic, elementary fermions provide Dirac partners for the composite fermion states filling up the rest of the right-handed top quark multiplet and can be searched for at future colliders.
\item \textbf{Gauge coupling unification:} The SM gauge couplings unify around the scale $10^{15}$~GeV due to the fact that the right-handed top quark is fully composite.  Proton decay is prevented by a baryon number symmetry respected by the strong and elementary sectors.  Preserving one-loop gauge coupling unification requires the exotic states to have masses below around 100~TeV, so the scale $f$ cannot be arbitrarily large.
\end{itemize}

\begin{figure}[!t]
\begin{center}
\includegraphics[width=0.35\textwidth]{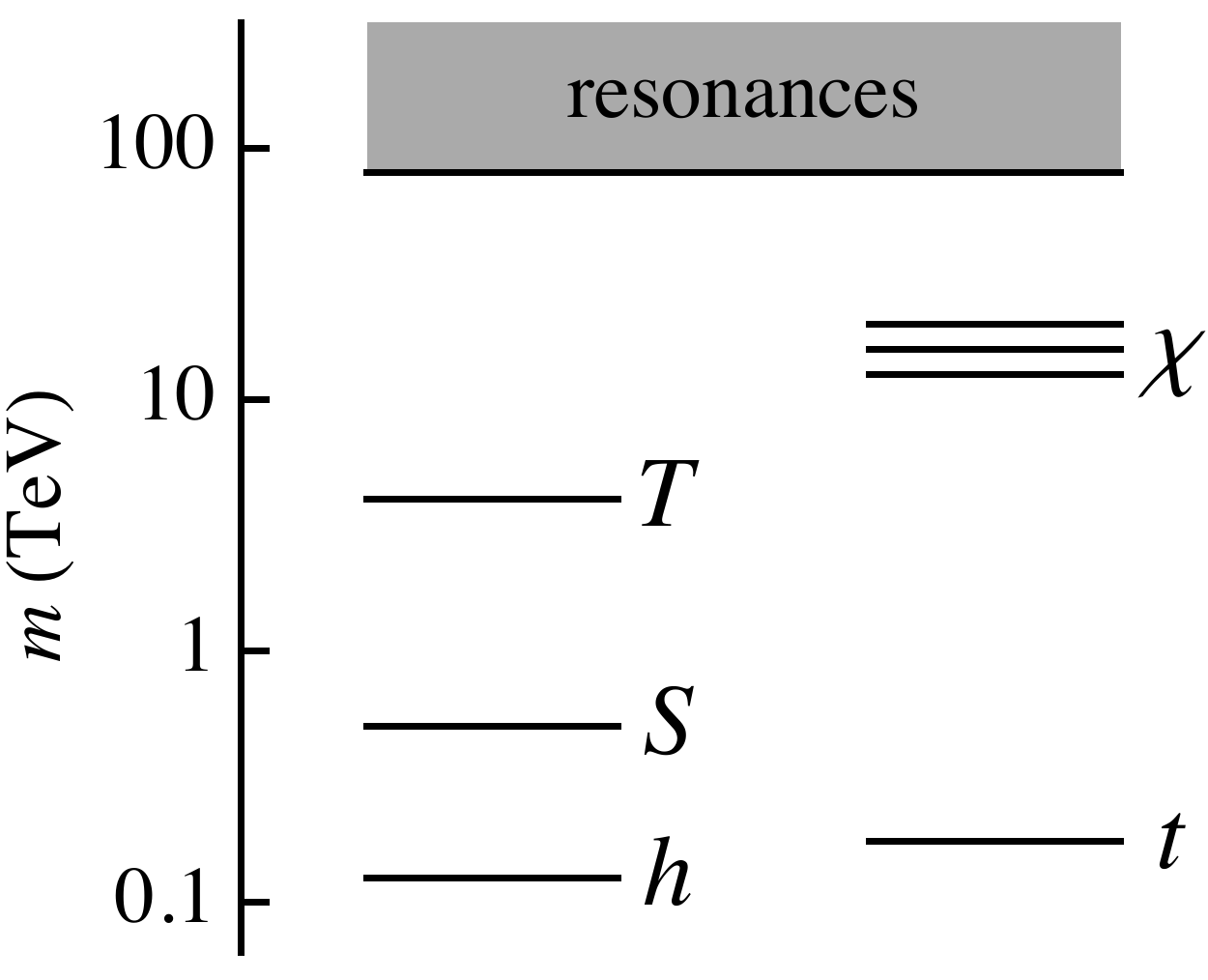}
\caption{A schematic diagram of the composite particle spectrum in the ``unnatural" composite Higgs model.  States on the left are bosons and those on the right are fermions\label{fig:spec}}
\end{center}
\end{figure}

The low energy spectrum is depicted in figure~\ref{fig:spec} for $f\sim10$~TeV\@.  It consists of resonance and exotic state masses in the 10 to 100~TeV range and pNGB masses in the 100~GeV to 1~TeV range. Some features in the ``unnatural" composite Higgs model have also recently been studied in the context of naturalness, these include scalar dark matter~\cite{Frigerio:2012uc} and grand unification~\cite{Frigerio:2011zg}. A connection between dark matter and unnaturalness was also considered in ref.~\cite{Vecchi:2013iza}.

The outline for the rest of this paper is as follows. In section~\ref{sec:GF}, we discuss the general features of composite Higgs models, including the precision-electroweak and flavour constraints, as well as the features needed to embed the models into a GUT\@.  The minimal coset space is presented in section~\ref{sec:MM}, and is chosen to incorporate both unification and a dark matter candidate.  The effective Lagrangian and pNGB potential is then derived and the pNGB masses are calculated.  Dark matter is discussed in section~\ref{sec:DM}, where the symmetry responsible for dark matter stability and the parameter space consistent with experiment is identified.  Section~\ref{sec:ES} is devoted to the phenomenology associated with decay of exotic states.  In particular, the colour-triplet pNGB is shown to be long-lived so that displaced decays at a collider are possible.  Our concluding remarks are given in section~\ref{sec:CO}.  The appendices contain the $U(7)$ representations and mathematical details used in deriving the pNGB potential.

\section{General features of composite Higgs models\label{sec:GF}}

Composite Higgs models are characterised by a strongly coupled sector with a global symmetry group, $G$, spontaneously broken at a scale $f$ to a subgroup, $H$.  This gives rise to a set of Nambu-Goldstone bosons living in the coset space $G/H$ and containing, possibly among other states, the Higgs doublet of the SM\@.  The strong sector itself is characterised by a set of operators, $\cal O$, and the masses of resonances emerging from the strong sector are of order $m_\rho = g_\rho f$, where $g_\rho$ is the generic coupling of the strong sector.

The usual framework for generating electroweak gauge interactions and Yukawa couplings in composite Higgs models is through that of partial compositeness.  That is, there is a mixing between the elementary and composite sectors which explicitly breaks the global symmetry of the strong sector.  The mixing is parameterised by
\be\label{eq:genmix}
\cL_{\rm mix}=\sum_\cA \cA{}_a^\mu\Omega{}_A^aT^A\cJ_\mu+\sum_\psi\psi_{i}(\lambda_\psi)^{i}_I(\cO_\psi)^I.
\ee

The first term describes the mixing of the elementary gauge fields, $\cA^\mu$, with the strong sector.  The index $a$ runs over all generators of the SM gauge group and $\Omega$ is a projector.  It projects the generators of the global symmetry in the strong sector, $T^A$, onto those of the SM gauge group.  It takes the explicit form
\be\label{eq:Omegadef}
\Omega{}_A^a=\nbrack{g_3\delta{}_A^{a_3}+g_2\delta{}_A^{a_2}+g_1\delta{}_A^{a_1}}
\ee
where $g_{1,2,3}$ are the SM gauge couplings and the $\delta$'s pick out the generators of the SM gauge group as they are embedded in $G$.  $\cJ_\mu$ is the $G$ current in the strong sector to which the elementary gauge fields couple.

The second term describes the mixing of the elementary fermions, $\psi$.  The index $i$ runs over the appropriate representation of the SM gauge group and the index $I$ runs over the representation of $G$ into which the elementary fermion is embedded.  The projector, $\lambda_\psi$, projects components of the strong sector operator, $\cO_\psi$, in a full $G$ representation, onto the SM representation containing $\psi$.  When they are permitted by the symmetries of the model these terms generate Yukawa couplings
\be
y_{\psi\psi^\prime}\sim\frac{|\lambda_\psi||\lambda_{\psi^\prime}|}{g_\rho}.
\ee

All of the projectors can be thought of as spurions parameterising the explicit breaking of the global symmetry in the strong sector.  When they are turned off the complete model has an enhanced global symmetry $G_{\rm SM}\times G$.  When they are turned on this is broken to a single, diagonal $G_{\rm SM}$ factor.  Hence they provide a convenient tool for keeping track of all explicit symmetry breaking~\cite{Mrazek:2011iu}.

\subsection{Precision-electroweak and flavour constraints}

The main precision-electroweak observables constraining composite Higgs models are the oblique parameters, $S$ and $T$.  There are two generic contributions to the $S$ parameter: one from the mixing with the $\rho$ mesons, leading to a contribution $\Delta S\propto M_W^2/m_\rho^2$, and the other from the deviation of the SM gauge field couplings $\Delta S\propto \alpha/(4\pi)\times v^2/f^2$.  These contributions result in an overall constraint of the form $f\gtrsim0.8$~TeV \cite{Contino:2010rs}.  However, it was pointed out in ref.~\cite{Bertuzzo:2012ya} that, in models based on a coset space with a non-flat geometry, there can be a contribution to the $T$ parameter at tree level unless it is forbidden by custodial symmetry.  The model we will consider corresponds to such a scenario and this tree level contribution, $\Delta T \propto v^2/f^2$, results in a stronger constraint, $f\gtrsim5.5$~TeV~\cite{Mrazek:2011iu}.

Flavour constraints also restrict the value of $f$.  We will assume that the flavour structure in the composite sector is anarchic and that all couplings are of order $g_\rho$.  The main constraint from flavour changing processes then comes from the Kaon observable, $\epsilon_{K}$, and leads to a lower bound $f\gtrsim10$~TeV \cite{Bellazzini:2014yua}.  Constraints from the B-meson sector give $f\gtrsim1$~TeV and constraints from the loop-driven dipole operator, which leads to flavour changing processes like $b\to s\gamma$ and CP violation in Kaons, give $f\gtrsim2$~TeV\@.  There are also constraints from the leptonic sector \cite{Agashe:2006iy,KerenZur:2012fr}, primarily from the tree-level $e\to\mu$ transition and loop-driven $\mu\to e\gamma$ processes.  These translate to a lower bound of $f\gtrsim10$~TeV\@. 

By choosing $f\gtrsim10$~TeV all constraints from precision-electroweak observables and the flavour sector are therefore comfortably satisfied.

\subsection{Top companions and unification}

Two features greatly improve SM gauge coupling unification in composite Higgs models.  The entire SM gauge group should be embedded into a simple subgroup of the strong sector's unbroken global symmetry, e.g.\ $SU(5)\subset H\subset G$, and the right-handed top quark should be fully composite.  The reasoning is as follows~\cite{Agashe:2005vg}.

At leading order, and neglecting threshold corrections, contributions to the running of the SM gauge couplings can be split into two components, one coming from the elementary sector and one from the strong sector.  Any SM states assumed to be fully composite should be subtracted from the elementary sector's contribution and included in the strong sector's contribution instead.  In a model with a composite Higgs and a fully-composite right-handed top quark this gives, schematically
\be
\alpha(\mu)=\mbox{SM}-\{H,t^c\}+\mbox{strong sector}+\mbox{elementary exotics}
\ee
allowing for elementary exotic states alongside the usual elementary SM degrees of freedom.

The embedding of the SM gauge group into a simple subgroup of the strong sector's unbroken global symmetry means that all composite objects, which necessarily respect the unbroken global symmetry, come in complete GUT multiplets.  Hence the strong sector does not contribute to the differential running.  This also means that a fully-composite right-handed top quark comes with other light, composite fermions corresponding to the rest of its GUT multiplet.  Since these fermions have not been observed, and since they would lead to SM gauge anomalies, we must add some exotic, elementary fermions to the model to provide them with Dirac partners.\footnote{Note that the exotic, elementary fermions do not come in a complete GUT multiplet, as the composite right-handed top must remain light.  Any states missing from the multiplet are assumed to get large masses from GUT-scale physics.  This is analogous to doublet-triplet splitting in the Higgs sector of supersymmetric GUTs.}  As far as the differential running is concerned this is equivalent to subtracting an additional conjugate right-handed top quark.  The overall effect is for the differential runnings to obey
\be\label{eq:diffalpha}
\alpha_i(\mu)-\alpha_j(\mu)=\mbox{SM}-\{H,t^c, \bar{t}^c\}
\ee
above all associated mass scales.  Equivalently, the one-loop beta-function coefficients, $b_i$ satisfy
\begin{align}
b_1-b_2 & =\frac{94}{15} & b_2-b_3 & =\frac{13}{3}
\end{align}
above all associated mass scales. This leads to a value of the ratio $(b_2-b_3)/(b_1-b_2) \simeq 0.69$, which is similar to the corresponding MSSM value 0.71.

To determine what the relevant mass scales are we can go into a little more detail by writing down a mixing term
\be
\cL\supset \chi \lambda_\chi\cO_t
\ee
between the exotic, elementary fermions, $\chi$, and the operator in the strong sector generating the right-handed top quark, $\cO_t$.  When the strong sector confines this generates a Dirac mass term with $m_\chi\sim\lambda_\chi f$, combining the elementary and composite Weyl fermions.  We will refer to the resulting mass eigenstates as top companions (not to be confused with top partners, which are massive resonances produced exclusively by the strong sector).  In addition, the mixing term explicitly breaks the global symmetry in the strong sector as the top companions do not come in complete representations of the GUT group; they are missing the right-handed top quark component.  This generates a mass for all of the pNGBs at one loop, of order $g_\rho\lambda_\chi f/(4\pi)\sim\lambda_\chi f$.  However, the models we will be considering will be tuned to keep the Higgs light so this mass only applies to any additional pNGBs in its GUT multiplet.  Eq.~\eqref{eq:diffalpha} therefore applies above the scale $m_\chi\sim\lambda_\chi f$.

\begin{figure}[!t]
\begin{center}
\includegraphics[height=4cm]{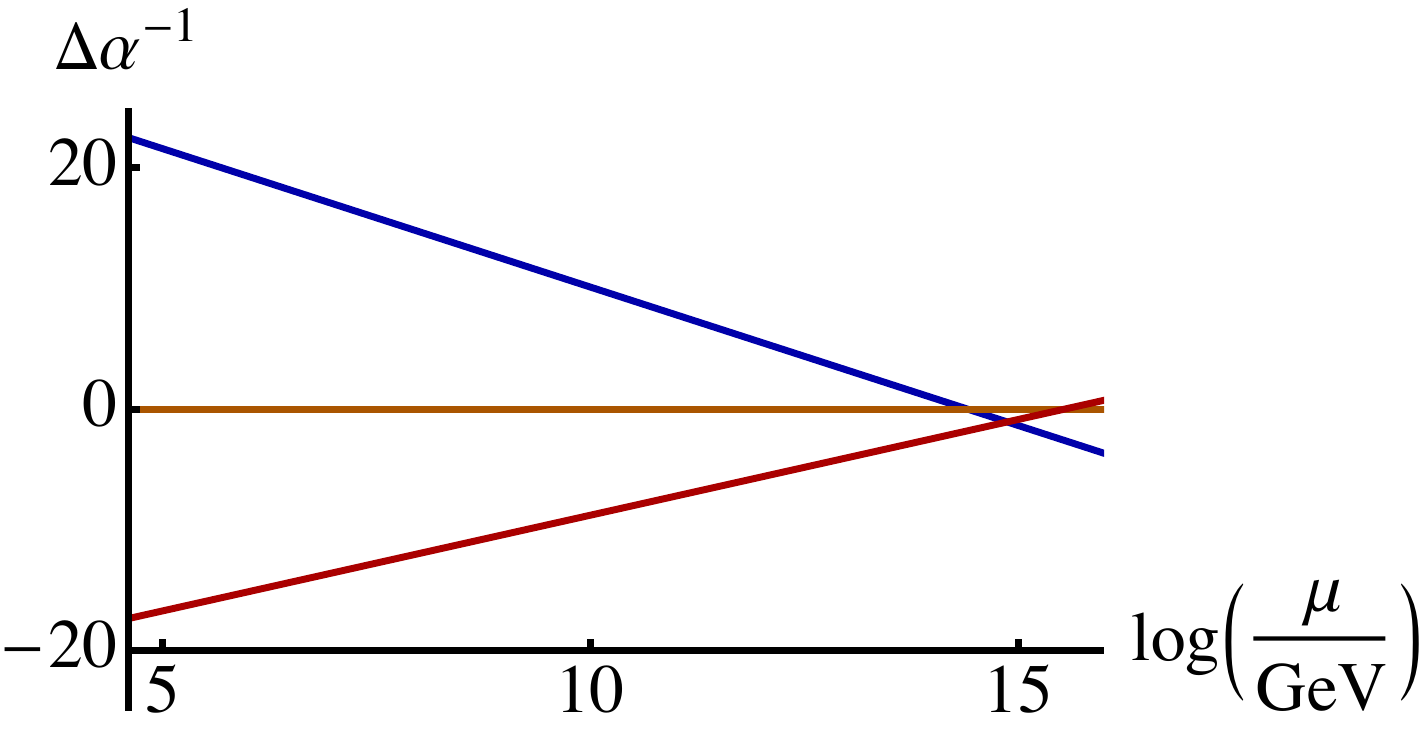}\hspace{0.5cm}
\includegraphics[height=4cm]{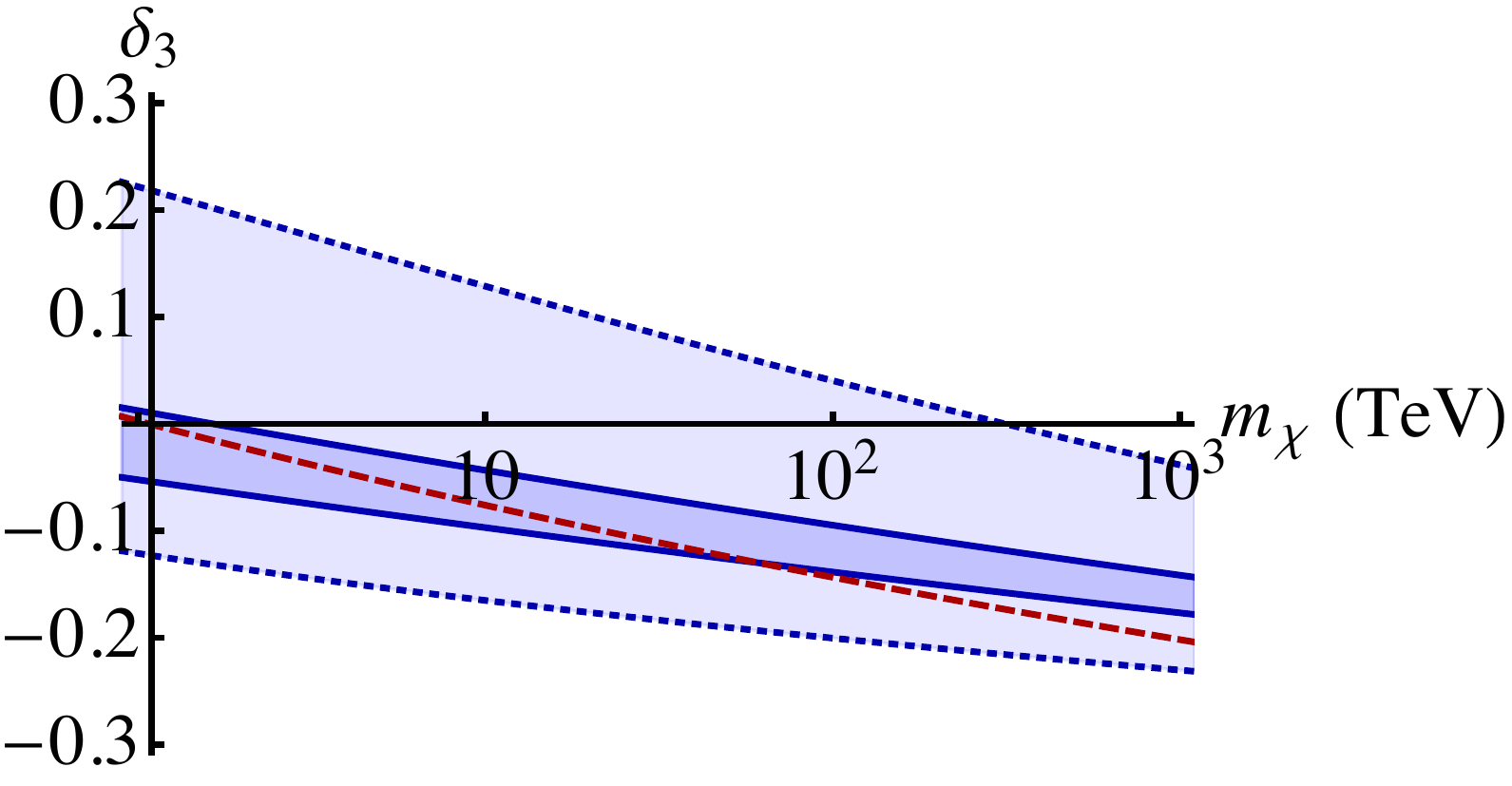}
\caption{{\em Left}: The running differential SM gauge couplings in a composite Higgs model with 20~TeV top companions.  $\Delta\alpha_i^{-1}\equiv\alpha_i^{-1}-\alpha_2^{-1}$.  {\em Right}: The postdiction of the QCD gauge coupling as a function of the top companion mass: $\delta_3\equiv[\alpha{}_3^{\rm th}(M_Z)-\alpha{}_3^{\rm ex}(M_Z)]/\alpha{}_3^{\rm ex}(M_Z)$.  The solid and dotted lines mark out the bands from the 4D calculation for $b_{\rm strong}=1,5$ respectively, the dashed line the result of the 5D calculation.\label{fig:run}}
\end{center}
\end{figure}

Choosing a top companion mass of 20~TeV one then finds greatly improved unification at leading order, as shown on the left of figure~\ref{fig:run}.  Since the unification scale is a little under $10^{15}$~GeV one may worry that composite GUTs are ruled out due to proton decays mediated by $X$ and $Y$ gauge bosons.  This is not the case, as we will discuss in the next subsection.

Going beyond leading order we can attempt to estimate the two-loop contributions to the running of the SM gauge couplings.  Those coming from the SM and top companions can be calculated exactly.  Other contributions from the strong sector are less certain.  Following closely the reasoning of ref.~\cite{Agashe:2005vg}, they can be parameterised by $b_{\rm strong}$, which is related to the number of colours in the strong sector.  Two-loop gauge contributions go like $B_{\rm strong}\sim9b_{\rm strong}$ and two-loop gauge-top-companion contributions go like $C\sim3\lambda_\chi b_{\rm strong}$. Including all of these contributions we have postdicted the QCD coupling and compared it to the observed value for a variety of top companion masses.  We find good agreement, as shown by the solid and dotted lines on the right of figure~\ref{fig:run}.  The solid lines bound the result for $b_{\rm strong}=1$, a band being given due to the theoretical uncertainty on the contribution from the strong sector, and the dotted lines bound the result for $b_{\rm strong}=5$.  Higher values of $b_{\rm strong}$ result in the gauge couplings diverging before unification.

We can also calculate the differential running of the SM gauge couplings in a 5D theory, then use the AdS/CFT correspondence to relate the results to the 4D models we consider here.  It should be noted that the correspondence cannot be completely trusted for these models as there is no guarantee that the strong sector is a large-$N$ gauge theory with a sufficiently large $N$.  After all, the limit on $b_{\rm strong}$ found above is not that large.

With this caveat in mind, the relevant 5D theory is a slice of AdS$_5$ with the Higgs identified as the fifth component of a 5D gauge field.  The SM fermions propagate in the bulk, with the right-handed top quark highly localised in the IR and the rest of the third generation more evenly distributed in the bulk.  Expressions for the running gauge couplings for this setup are given in ref.~\cite{Choi:2002ps,Gherghetta:2004sq}.  Postdicting the QCD gauge couplings as before, we find good agreement with both the observed value and our previous estimate.  This is shown by the dashed line on the right of figure~\ref{fig:run}.

The upper bound this argument imposes on the top companion mass, and therefore the symmetry breaking scale, $f$, is obviously subjective.  Things are further complicated by the uncertainty inherent to the two-loop contributions.  However, requiring the outer band shown in figure~\ref{fig:run} (from the 4D calculation with $b_{\rm strong}=5$) to contain $\delta_3=0$ leads to an upper bound on $f$ of between 100~TeV and 1000~TeV, assuming $\lambda_\chi$ is of order one.  This corresponds to an accuracy of between 10\% and 20\% using the inner band or the 5D calculation.  For higher values of $f$ none of the bands contain $\delta_3=0$.  Precision unification would then need to rely on large corrections coming from some other sector or on even higher-order effects.

\subsection{Baryon and lepton number}

As with any GUT framework there is a risk in composite GUTs of inducing baryon- and lepton-number-violating operators that are in conflict with observations.  Indeed, this is well known to be the case in supersymmetric GUTs (or even just in the MSSM) where the solution is to impose an additional global symmetry, R-parity, by hand.  In composite GUTs a completely analogous approach is taken, the simplest option being to assume that both baryon and lepton number are respected by the strong sector \cite{Agashe:2004ci, Agashe:2004bm}.

If baryon number was not respected by the strong sector one would generically expect baryon-number-violating operators to be generated, suppressed only by the compositeness scale rather than the much higher GUT scale.  These terms would arise due to resonance exchange in the strong sector, and would lead to a proton lifetime far shorter than the observed value.  Similarly, if lepton number was not respected by the strong sector, the strong sector would generically generate a Majorana neutrino mass term suppressed only by the compositeness scale, thus leading to neutrino masses larger than those observed.

Proton stability is further ensured in composite GUTs as light matter states do not necessarily come in complete GUT multiplets~\cite{Agashe:2004ci, Agashe:2004bm}.  For example, in minimal $SU(5)$ the quark doublet lives in a different ${\bf10}$ to the charged lepton singlet; the other states in each multiplet are assumed to get large masses from GUT-scale physics (as has already been seen for the top companions).  When the mixing between the strong and elementary sectors also respects baryon number, as will be the case in our model, all GUT multiplets then have a well-defined baryon number and the only potential sources of baryon number violation are sphaleron processes and gravitational effects.  Proton decays mediated by $X$ and $Y$ gauge bosons, or by the Higgs colour-triplet states expected in many GUTs, are simply forbidden.  Practically, this is because the `leptons' appearing in any problematic operators are no longer light, SM states.  Instead, they are exotic, GUT-scale fermions with non-SM-like baryon numbers.

\section{$SU(7)/SU(6)\times U(1)$: the minimal model\label{sec:MM}}

If we want pNGB dark matter the coset space of our model must contain a SM singlet stabilised by an unbroken global symmetry.  The smallest coset space that contains $SU(5)$ and has enough pNGBs to accommodate both a Higgs multiplet and a SM singlet is $SU(6)/SU(5)$.  This contains eleven pNGBs: a complex ${\bf 5}$ of $SU(5)$, $\tilde{H}$, and a real singlet, $\tilde{S}$.  Since the singlet is real the only global symmetry that could stabilise it is a $Z_2$ symmetry.  The only such symmetry acting on the pNGBs is charge conjugation, which we know is ultimately broken.  Hence $SU(6)/SU(5)$ cannot work.

Another possibility is $U(6)/U(5)$.  The pNGBs are the same as before but there is now an additional $Z_2$ symmetry that is unrelated to charge conjugation.  Nonetheless this symmetry does not act on the pNGBs in the correct way; it maps $\tilde{H}\to-\tilde{H}$ and $\tilde{S}\to\tilde{S}$ and is anyway broken by the VEV responsible for breaking $U(6)$ to $U(5)$.  We are therefore forced to move to a larger coset space, the simplest option being $SU(7)/SU(6)\times U(1)$.  Other coset spaces such as $SO(12)/SO(11)$ (which also includes a custodial symmetry) provide additional possibilities, although generically lead to models with more exotic fermions.

It is worth pointing out that this coset space has the additional advantage of being straightforward to realise in a theory of gauge fields and fermions only, as the symmetry breaking is due to a single spurion in the adjoint representation of the $SU(7)$ symmetry \cite{Barnard:2013zea}.  There are therefore good prospects for a UV completion of the model we will present here.

\subsection{The pNGB sector}

There are twelve pNGBs in the $SU(7)/SU(6)\times U(1)$ coset space: a complex ${\bf5}$ of $SU(5)\subset SU(6)$, $\tilde{H}$, and a complex $SU(5)$ singlet, $\tilde{S}$.  They can be collected into a ${\bf7}$ of $SU(7)$ as
\be\label{eq:Sigma7p}
w=e^{i\Pi}\begin{pmatrix}0_{(6)}\\1\end{pmatrix}=
\frac{1}{\sqrt{|\tilde{H}|^2+|\tilde{S}|^2}}\begin{pmatrix}i\tilde{H}\sin\pfrac{\sqrt{|\tilde{H}|^2+|\tilde{S}|^2}}{f}\\i\tilde{S}\sin\pfrac{\sqrt{|\tilde{H}|^2+|\tilde{S}|^2}}{f}\\\sqrt{|\tilde{H}|^2+|\tilde{S}|^2}\cos\pfrac{\sqrt{|\tilde{H}|^2+|\tilde{S}|^2}}{f}\end{pmatrix}
\ee
where $\Pi$ is the pNGB matrix, divided by the decay constant, $f$.  It will be more convenient to work in terms of a gauge basis in which
\be\label{eq:Sigma7}
w=\frac{1}{f}\begin{pmatrix}H\\S\\\sqrt{f^2-|H|^2-|S|^2}\end{pmatrix}.
\ee
There is one spurion, a ${\bf48}$ (adjoint) of $SU(7)$, that can be used to realise the symmetry breaking pattern.  This is parametrised by $w$ as
\be
{\bf48}=\frac{1}{\sqrt{42}}\nbrack{7ww^\dag-\uno},
\ee
the normalisation being chosen such that $\tr{{\bf48}^2}=1$.  The component proportional to the identity matrix does not contribute any new terms to the effective Lagrangian so, without loss of generality, we can consider the simplified spurion
\be\label{eq:Sigmau}
\Sigma=ww^\dag.
\ee
Because $w^\dag w=1$ the number of independent projectors that can be formed from $w$ is finite.

The complete pNGB embeddings are given by table~\ref{tab:pNGB}.  The unbroken $SU(6)\times U(1)$ global symmetry contains two $U(1)$ subgroups that commute with its $SU(5)$ subgroup;  $U(1)_7$ is the $U(1)$ symmetry contained in $SU(7)$ but not $SU(6)$, and $U(1)_6$ is the $U(1)$ symmetry contained in $SU(6)$ but not $SU(5)$.  Of particular interest in the linear combination
\be\label{eq:U1def}
U(1)_S\equiv\frac{1}{42}\sbrack{U(1)_7-7U(1)_6}
\ee
under which only the singlet is charged.  Hence $U(1)_S$ is not broken by the Higgs VEV, whereupon the singlet may be stabilised and can provide a dark matter candidate.

Although $U(1)_S$ provides the foundation for the symmetry stabilising the dark matter candidate in this model it will turn out that some light, elementary fermions carry charge under it.  The situation is then a little more complicated.  However, enlarging $SU(7)$ to $U(7)=SU(7)\times U(1)_E$ and allowing $U(1)_S$ to mix with $U(1)_E$ and baryon number is a simple solution.  Since the adjoint of $SU(7)$ is neutral under $U(1)_E$ and baryon number these symmetries are spectators to the spontaneous symmetry breaking and the pNGB sector remains as described.

\begin{table}[!t]
\be\nonumber
\begin{array}{|l|r|rr|rr|}\hline
& SU(7) & SU(6) & U(1)_7 & SU(5) & U(1)_S \\\hline
H &\multirow{2}{*}{\bf48} & \multirow{2}{*}{\bf6}& \multirow{2}{*}{7} & {\bf5} & 0 \\
S &&&& {\bf1} & 1 \\\hline\end{array}
\ee
\caption{The pNGB embeddings in the $SU(7)/SU(6)\times U(1)$ model.  $U(1)_S$ is defined in eq.~\eqref{eq:U1def} and the charges under $U(1)_7$ and $U(1)_6$ are given in table~\ref{tab:SU7}.\label{tab:pNGB}}
\end{table}

\subsection{Matter embeddings}

The elementary SM fermions, $(q,l,u^c,d^c,e^c)$, come in $SU(3)\times SU(2)\times U(1)_Y$ multiplets and are embedded into $SU(5)\times U(1)_L\times U(1)_B$ multiplets as follows:
\begin{align}
q & \in{\bf10}_{0,\frac{1}{3}} & u^c & \in{\bf10}_{0,-\frac{1}{3}} & d^c & \in{\bf\overline{5}}_{0,-\frac{1}{3}} & e^c & \in{\bf10}_{-1,0} & l & \in{\bf\overline{5}}_{1,0}.
\end{align}
As is usual in composite GUTs, and as has already been seen for the top companions, the elementary fermions do not fill out complete representations of the GUT group.  Instead, the remaining components are assumed to acquire GUT-scale masses due to high-scale physics, analogously to doublet-triplet splitting in supersymmetric GUTs.

When considering which representations of $SU(7)$ to embed the elementary fermions into we need to ensure that our choice allows for all quark and lepton Yukawa couplings.  Since these are generated through mixing with the strong sector it must be possible to write down terms that contain the desired Yukawa couplings and respect the unbroken $SU(6)\times U(1)_7$ symmetry of the strong sector.

Table~\ref{tab:SU7} lists the decompositions of several of the smaller representations of $SU(7)$ and will be helpful throughout this section.

\subsubsection{Quarks}

Up-type Yukawa couplings have $SU(5)$ structure ${\bf10}(q){\bf10}({u^c}){\bf5}(H)$ so we need representations of $SU(7)$, ${\bf R}_7(q)$ and ${\bf R}_7(u^c)$, such that both ${\bf R}_7(q)$ and ${\bf R}_7(u^c)$ contain a ${\bf10}$ of $SU(5)$, and ${\bf R}_7(q)\times{\bf R}_7(u^c)$ contains a singlet of $SU(6)\times U(1)_7$.  The unique choice among the smaller representations is to have one of ${\bf R}_7(q)$ and ${\bf R}_7(u^c)$ equal to ${\bf35}$, the three-index antisymmetric representation of $SU(7)$, and the other equal to $\overline{\bf35}$.  Down-type Yukawa couplings have $SU(5)$ structure ${\bf10}(q)\overline{\bf5}(d^c)\overline{\bf5}(H)$ so we also need representations of $SU(7)$, ${\bf R}_7(q)$ and ${\bf R}_7(d^c)$, such that ${\bf R}_7(q)$ contains a ${\bf10}$ of $SU(5)$, ${\bf R}_7(d^c)$ contains a $\overline{\bf5}$ of $SU(5)$, and ${\bf R}_7(q)\times{\bf R}_7(d^c)$ contains a singlet of $SU(6)\times U(1)_7$.  Choosing ${\bf R}_7(q)$ equal to ${\bf35}$ and ${\bf R}_7(d^c)$ equal to $\overline{\bf35}$ meets this requirement.

The quark embeddings given in table~\ref{tab:matt} provide a simple choice consistent with all Yukawa couplings.  Note that the left-handed quark doublet mixes with two different strong sector operators, each of which acts as the Dirac partner to one of $\cO_{u^c}$ and $\cO_{d^c}$.  Two operators are needed because the embeddings of the left-handed quark doublet required to generate both Yukawa couplings would otherwise violate the symmetry that stabilises the dark matter candidate in this model.  This is discussed in more detail in section~\ref{sec:DM}.

\subsubsection{Leptons}

Charged-lepton Yukawa couplings have the same $SU(5)$ structure as down-type Yukawa couplings.  This time we will choose an embedding with ${\bf R}_7(e^c)$ equal to ${\bf21}$, the two-index antisymmetric representation of $SU(7)$, and ${\bf R}_7(l)$ equal to $\overline{\bf21}$.  Assuming the existence of a right-handed neutrino, $N^c$, the neutrino Yukawa couplings have $SU(5)$ structure $\overline{\bf5}(l){\bf1}(N^c){\bf5}(H)$.  This can be generated by choosing ${\bf R}_7(l)$ to be equal to $\overline{\bf21}$ as before and ${\bf R}_7(N^c)$ equal to ${\bf21}$.  Again, and as will be discussed in more detail later, the left-handed lepton doublet must couple to two different operators in the strong sector if we want to preserve the symmetry stabilising the dark matter candidate.

\begin{table}[!t]
\be\nonumber
\begin{array}{|l|r|r|rrr|}\hline
& SU(7) & SU(6) & SU(5) & U(1)_L & U(1)_B \\\hline
q_{(u)} & \overline{\bf35} & {\bf20} & {\bf10} & 0 & \frac{1}{3} \\
q_{(d)} & {\bf35} & {\bf20} & {\bf10} & 0 & \frac{1}{3} \\
u^c & {\bf35} & {\bf15} & {\bf10} & 0 & -\frac{1}{3} \\
d^c & \overline{\bf35} & \overline{\bf15} & \overline{\bf5} & 0 & -\frac{1}{3} \\
l_{(\nu)} & \overline{\bf21} & \overline{\bf15} & \overline{\bf5} & 1 & 0 \\
l_{(e)} & \overline{\bf21} & \overline{\bf6} & \overline{\bf5} & 1 & 0 \\
N^c & {\bf21} & {\bf6} & {\bf1} & -1 & 0 \\
e^c & {\bf21} & {\bf15} & {\bf10} & -1 & 0 \\
(\tilde{q}^c,\tilde{e}) & \multirow{2}{*}{$\overline{\bf35}$} & \multirow{2}{*}{$\overline{\bf15}$} & \overline{\bf10} & 0 & \frac{1}{3} \\
(\tilde{d}^c,\tilde{l}) &&& \overline{\bf5} & 0 & 0 \\\hline\end{array}
\ee
\caption{Elementary fermion embeddings.  The decompositions of the $SU(7)$ representations are detailed in table~\ref{tab:SU7}.  The subscripts on $q_{(u)}$ and $q_{(d)}$ denote the embeddings responsible for generating the up- and down-type Yukawas respectively, and similarly in the lepton sector for $l_{(\nu)}$ and $l_{(e)}$.\label{tab:matt}}
\end{table}

\subsubsection{Top companions}

The strong sector operator generating the right-handed top quark decomposes to an $SU(5)$ representation through the chain ${\bf35}\ni{\bf15}\ni{\bf10}$.  The elementary fermions, $\chi$, that give masses to the rest of the right-handed top quark multiplet, must come in a complete conjugate representation of $SU(6)\times U(1)_E\times U(1)_7$, sans a conjugate right-handed top quark, i.e.\
\be
\chi\in\overline{\bf15}_{-3,4}
\ee
such that all unwanted fermions get masses.  Masses in the strong sector respect the full $SU(6)\times U(1)_E\times U(1)_7$ 
symmetry, so filling up $SU(5)$ multiplets alone is not enough.  Breaking this down into SM-like degrees of freedom under 
$SU(3)\times SU(2)\times U(1)_Y$, we find top companions
\be
\chi \equiv\tilde{q}^c\oplus\tilde{e}\oplus\tilde{d}^c\oplus\tilde{l}=
(\overline{\bf3},{\bf2})_{-\frac{1}{6}}\oplus({\bf1},{\bf1})_{-1}\oplus(\overline{\bf3},{\bf1})_{\frac{1}{3}}\oplus({\bf1},{\bf2})_{-\frac{1}{2}}
\ee
where the multiplets are assembled as in table~\ref{tab:matt}.  One may notice that the different top companion $SU(5)$ multiplets end up carrying different baryon number.  This is because the true baryon number in this model ends up being a mixture of an external baryon number and some of the $U(1)$ symmetries contained within the unbroken global symmetry in the strong sector.  This is discussed in more detail in section~\ref{sec:DM}.

\subsection{The effective Lagrangian and pNGB potential}

The mixing Lagrangian for the third generation of matter in its full glory is
\begin{align}\label{eq:Lmix}
\cL\supset{} & {}(\tilde{q}^c,\tilde{e})\lambda{}_\chi^{\bf10}\cO{}_t^{\bf35}+(\tilde{d}^c,\tilde{l})\lambda{}_\chi^{\bf5}\cO{}_t^{\bf35}+q\lambda_t\cO{}_t^{\bf35}+q\lambda_b\cO{}_b^{\overline{\bf35}}+b^c\lambda_{b^c}\cO{}_{b^c}^{\bf35}{}\nonumber\\
& +{}l\lambda_\nu\cO{}_\nu^{\bf21}+l\lambda_\tau\cO{}_\tau^{\bf21}+N^c\lambda_{N^c}\cO{}_{N^c}^{\overline{\bf21}}+\tau^c\lambda_{\tau^c}\cO{}_{\tau^c}^{\overline{\bf21}}+m_NN^cN^c
\end{align}
where the projectors are defined explicitly in appendix~\ref{app:lambda}\@.  The left-handed quark doublet mixes with two different strong sector operators.  The coupling to the operator $\cO{}_t^{\bf35}$ generates the top quark Yukawa, $y_t\sim|\lambda_t|$, whereas the coupling to $\cO{}_b^{\overline{\bf35}}$ takes part in generating the smaller bottom quark Yukawa, $y_b\sim|\lambda_b||\lambda_{b^c}|/g_\rho$.  This story repeats in the lepton sector, but with the addition of a Majorana mass, $m_N$, for the right-handed neutrino.

The quark doublet, right-handed bottom quark and exotic, elementary fermions are the only elementary fermions expected to make a significant contribution to the pNGB potential.  Upon integrating out the strong sector, and using the above mixing Lagrangian, the relevant terms in the effective Lagrangian containing the pNGBs are
\begin{align}\label{eq:Lefff}
\cL_{\rm eff}\supset{} & (\bar{\tilde{q}}^c,\bar{\tilde{e}})_{i_4i_2}\slashed{p}(\tilde{q}^c,\tilde{e})^{j_4j_2}\sbrack{\Pi^{\chi\chi}(\lambda{}_\chi^{{\bf10}*})^{i_4i_2}_{IJK}(\lambda{}_\chi^{\bf10})_{j_4j_2}^{IJL}}\Sigma{}^K_L{} \nonumber\\
& +(\bar{\tilde{q}}^c,\bar{\tilde{e}})_{i_4i_2}\slashed{p}(\tilde{d}^c,\tilde{l})^{j_5}\sbrack{\Pi^{\chi\chi}(\lambda{}_\chi^{{\bf10}*})^{i_4i_2}_{IJK}(\lambda{}_\chi^{\bf5})_{j_5}^{IJL}}\Sigma{}^K_L+\mbox{h.c.}{} \nonumber\\
& +(\bar{\tilde{d}}^c,\bar{\tilde{l}})_{i_5}\slashed{p}(\tilde{d}^c,\tilde{l})^{j_5}\sbrack{\Pi^{\chi\chi}(\lambda{}_\chi^{{\bf5}*})^{i_5}_{IJK}(\lambda{}_\chi^{\bf5})_{j_5}^{IJL}}\Sigma{}^K_L{} \nonumber\\
& +\bar{q}^{i_3i_2}\slashed{p}q_{j_3j_2}\sbrack{\Pi^{tt}(\lambda_t^*)_{i_3i_2,IJK}(\lambda_t)^{j_3j_2,IJL}+\Pi^{bb}(\lambda_b^*)_{i_3i_2}^{IJL}(\lambda_b)^{j_3j_2}_{IJK}}\Sigma{}^K_L{} \nonumber\\
& +\bar{b}^c_{i_3}\slashed{p}b^{cj_3}\sbrack{\Pi^{b^cb^c}(\lambda_{b^c}{}^*)^{i_3}_{IJK}(\lambda_{b^c})_{j_3}^{IJL}}\Sigma{}^K_L{} \nonumber\\
& +(\bar{\tilde{q}}^c,\bar{\tilde{e}})_{i_4i_2}\slashed{p}q_{j_3j_2}\sbrack{\Pi^{\chi t}(\lambda{}_\chi^{{\bf10}*})^{i_4i_2}_{IJK}(\lambda_t)^{j_3j_2,IJL}}\Sigma{}^K_L+\mbox{h.c.}{} \nonumber\\
& +(\bar{\tilde{d}}^c,\bar{\tilde{l}})_{i_5}\slashed{p}q_{j_3j_2}\sbrack{\Pi^{\chi t}(\lambda{}_\chi^{{\bf5}*})^{i_5}_{IJK}(\lambda_t)^{j_3j_2,IJL}}\Sigma{}^K_L+\mbox{h.c.}{} \nonumber\\
& +q_{i_3i_2}b^{cj_3}\sbrack{M^{bb^c}(\lambda_b)^{i_3i_2}_{IJK}(\lambda_{b^c})_{j_3}^{IJL}}\Sigma{}^K_L+\mbox{h.c.}
\end{align}
where the $\Pi$'s and $M^{bb^c}$ are momentum-dependent form factors encoding the details of the strong sector.  It is readily checked that these are the only independent terms one can write down for these fermions at quadratic order in the projectors defined in appendix~\ref{app:lambda} and the spurion defined in eq.~\eqref{eq:Sigmau}.

\begin{figure}[!t]
\begin{center}
\includegraphics[height=2.1cm]{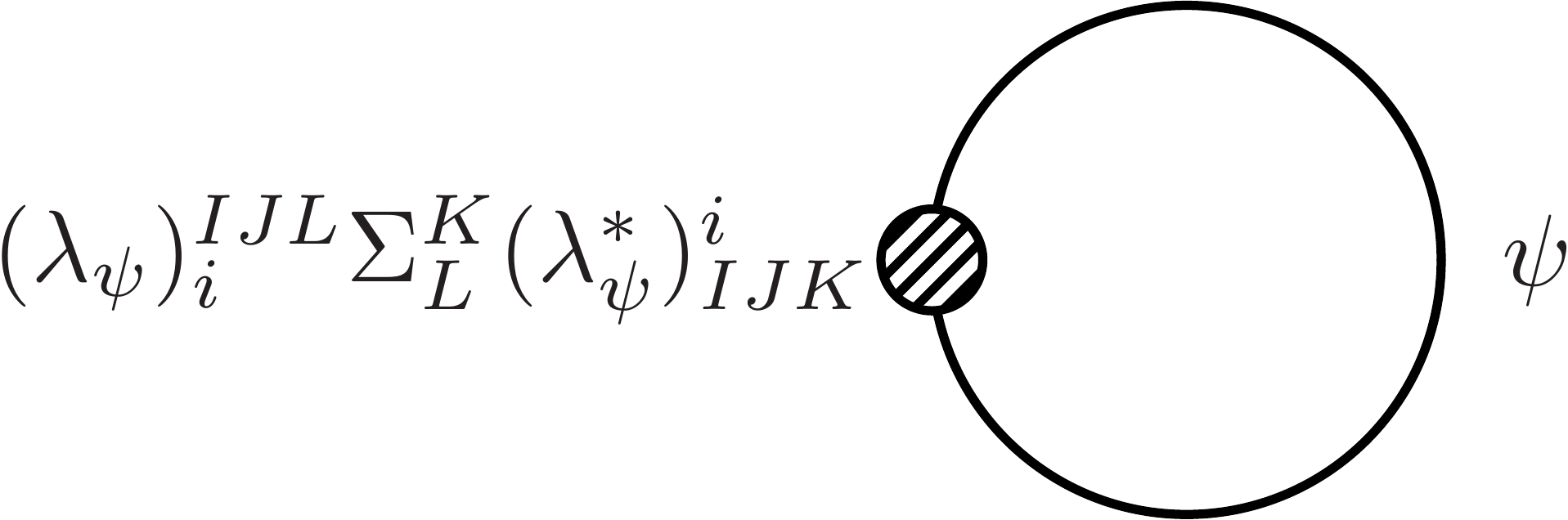}
\caption{Elementary fermion loops that generate leading order contributions to the pNGB potential.\label{fig:lambda2}}
\end{center}
\end{figure}

From eq.~\eqref{eq:Lefff}, and using eqs.~\eqref{eq:Sigma7} and \eqref{eq:Sigmau} to substitute in the components of the spurion, $\Sigma$, we can now calculate the matter contribution to the pNGB potential.  At one loop in the elementary fermions and at leading (quadratic) order in the $\lambda$'s this contribution comes from diagrams like the one shown in figure~\ref{fig:lambda2}.  The result is
\begin{align}\label{eq:Vmatter}
V_{\rm matter}={} & \frac{g_\rho^2f^2}{24\pi^2}c{}_1^\chi|\lambda_\chi|^2\nbrack{12-9|T|^2-7|D|^2-7|S|^2}+\frac{g_\rho^2f^2}{24\pi^2}c{}_1^t|\lambda_t|^2\nbrack{4|T|^2+3|D|^2}{} \nonumber\\
&+{}\frac{g_\rho^2f^2}{24\pi^2}c{}_1^b|\lambda_b|^2\nbrack{2|T|^2+3|D|^2+6|S|^2}+\frac{g_\rho^2f^2}{24\pi^2}c{}_1^{b^c}|\lambda_{b^c}|^2\nbrack{3-2|T|^2-3|D|^2}.
\end{align}
where the $c$'s are unknown, order-one coefficients.  Full details of this calculation can be found in appendix~\ref{app:VpNGB}.

\begin{figure}[!t]
\begin{center}
\includegraphics[height=2.1cm]{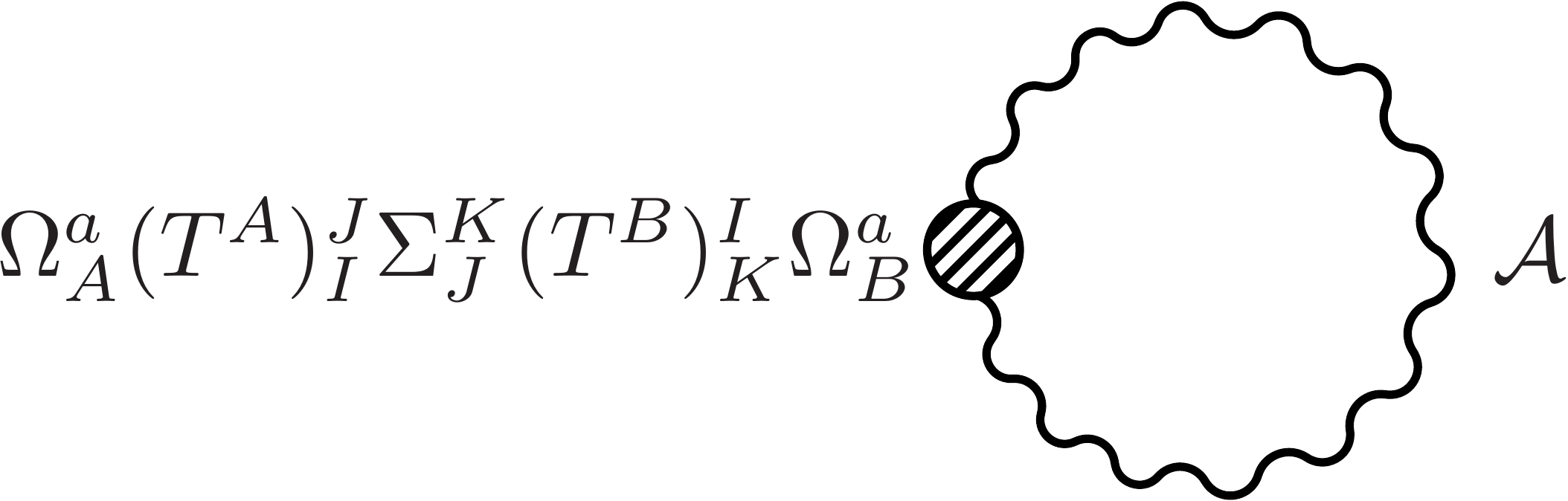}\hspace{1cm}
\includegraphics[height=2.1cm]{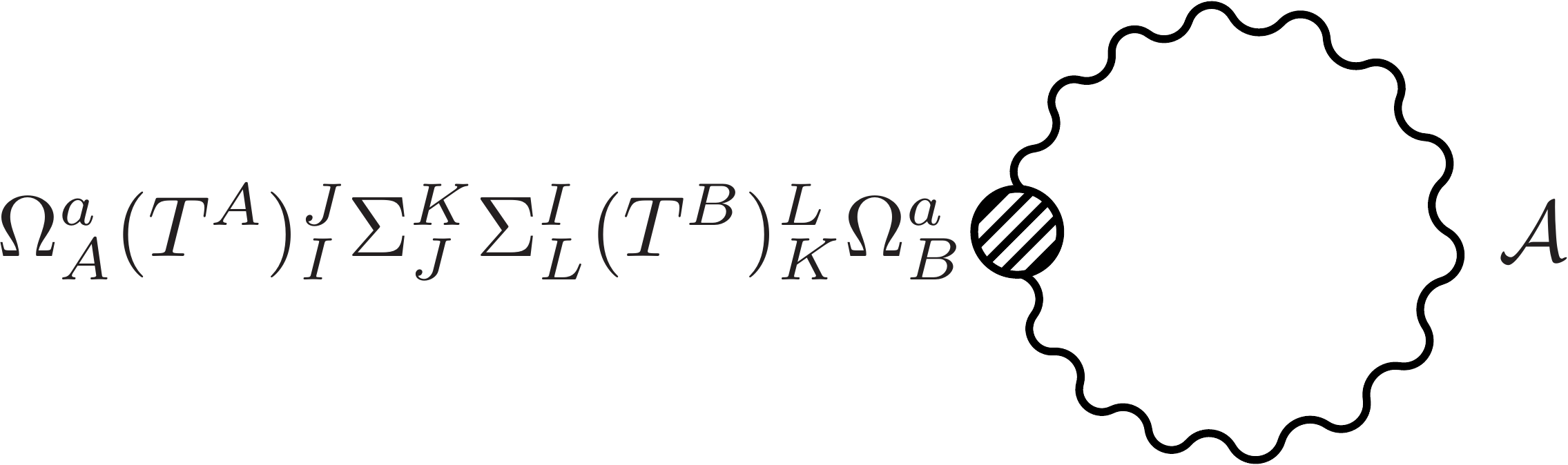}
\caption{Elementary gauge field loops that generate leading order contributions to the pNGB potential.\label{fig:omega2}}
\end{center}
\end{figure}

Elementary gauge fields also contribute to the pNGB potential.  The relevant terms in the effective Lagrangian are
\be\label{eq:Leffg}
\cL_{\rm eff}\supset\frac{1}{2}\cA{}_a^\mu P{}^T_{\mu\nu}\cA{}_b^\nu\sbrack{\Pi{}_1^{\rm A}\Omega{}_A^a(T^A)_I^J\Sigma_J^K(T^B)_K^I\Omega{}_B^b+\Pi{}_2^{\rm A}\Omega{}_A^a(T^A)_I^J\Sigma_J^K(T^B)_K^L\Sigma_L^I\Omega{}_B^b}
\ee
with the projector, $\Omega$, defined in eq.~\eqref{eq:Omegadef} and where $P{}^T_{\mu\nu}\equiv g_{\mu\nu}-p_\mu p_\nu/p^2$ is the transverse projection operator.  At one loop in the elementary gauge fields and at leading (quadratic) order in the $\Omega$'s the contribution comes from diagrams like the ones shown in figure~\ref{fig:omega2}.  The result is
\be\label{eq:Vgauge}
V_{\rm gauge}=\frac{3g_\rho^2f^2}{16\pi^2}c{}_1^{\rm A}\nbrack{\frac{4}{3}g_3^2|T|^2+\frac{3}{4}g_2^2|D|^2}+\frac{3g_\rho^2f^2}{16\pi^2}c{}_2^{\rm A}\nbrack{\frac{1}{3}g_3^2|T|^4+\frac{1}{4}g_2^2|D|^4}.
\ee
Again, the $c$'s are unknown, order-one coefficients and full details of this calculation can be found in appendix~\ref{app:VpNGB}.
While the $c$ coefficients can be of either sign in our effective, low energy framework, the sign of $c_1^A$ is positive in known calculable examples~\cite{Contino:2003ve}.

Putting everything together we can check whether the origin is unstable (so that electroweak symmetry is broken) and whether a suitable vacuum in which $|D|=v/f$ and $|T|=|S|=0$ (so that colour and the symmetry stabilising $S$ are both unbroken) exists.  Assuming that this is the case the potential in the doublet direction is
\be
V(|D|)=-\frac{\alpha}{f^2}|D|^2+\frac{\beta}{f^4}|D|^4
\ee
where
\begin{align}\label{eq:couplings}
\alpha & =\frac{g_\rho^2}{16\pi^2}f^4\nbrack{\frac{14}{3}c{}_1^\chi|\lambda_\chi|^2-2c{}_1^t|\lambda_t|^2-2c{}_1^b|\lambda_b|^2+2c{}_1^{b^c}|\lambda_{b^c}|^2-\frac{9}{4}c{}_1^{\rm A}g_2^2} \nonumber\\
\beta & =\frac{g_\rho^2}{16\pi^2}f^4\nbrack{\frac{3}{4}c{}_2^{\rm A}g_2^2}.
\end{align}
Electroweak symmetry breaking occurs whenever $\alpha>0$ and $\beta>0$ with a Higgs VEV
\be
v=f\sqrt{\frac{\alpha}{\beta}}.
\ee
The source of tuning is immediately apparent.  One generically finds $\beta\sim\alpha$ and so $v\sim f$ unless the parameters in the expression for $\alpha$ are tuned.

The mass of the physical Higgs boson in the electroweak symmetry breaking vacuum is given by
\be\label{eq:mh}
m_h^2=\frac{2\beta v^2}{f^4}=\frac{3c{}_2^{\rm A}g_\rho^2}{8\pi^2}M_W^2,
\ee
where $M_W = g_2v/2$.  Substituting in the observed values for the $W$-boson and Higgs masses this expression implies that $c{}_2^{\rm A}\sim64/g_\rho^2$.  The expected range for the strong sector coupling $4\lesssim g_\rho\lesssim4\pi$ is therefore completely consistent with $c{}_2^{\rm A}$ being an order-one coefficient; we cannot have $g_\rho\lesssim4$ as we need $|\lambda_t|\sim1$ (for the top quark Yukawa) and $|\lambda|/g_\rho\ll1$ (so that the $\lambda$'s can be considered as perturbations to the strong sector).

An interesting feature of this model is that the Higgs mass is proportional to the $W$-boson mass at leading order in the projectors; it is not proportional to the top quark or top partner masses as is the case in many composite Higgs models.  There are two reasons for this.  First, the geometry of the coset space we use is not flat so there is a contribution to the quartic coupling \eqref{eq:couplings} proportional to $g_2^2$, much like the $D$-term contribution one finds in supersymmetric models~\cite{Bertuzzo:2012ya}.  This term is not present in, for example, the minimal $SO(5)/SO(4)$ coset space.  Second, the $SU(7)$ representations we have used for the top quark do not allow for a leading order contribution to the quartic coupling.  This also happens in the minimal $SO(5)/SO(4)$ model for certain choices of top quark representation~\cite{Marzocca:2012zn, Pomarol:2012qf}.

For larger values of the unknown, order-one coefficients the next-to-leading order elementary fermion contributions (quartic in the $\lambda$'s) can, potentially, compete with the leading order term.  Nonetheless, they remain similar in magnitude to the gauge field contribution so the expected value for Higgs mass is broadly unchanged. Furthermore, in models where $c{}_2^{\rm A}$ is negative the next-to-leading order contributions can also ensure that the Higgs mass-squared remains positive.

The masses of the scalar triplet and singlet in the above vacuum are given by
\begin{align}\label{eq:TSmasses}
m_T^2 & \approx\frac{g_\rho^2}{16\pi^2}f^2\nbrack{-6c{}_1^\chi|\lambda_\chi|^2+\frac{8}{3}c{}_1^t|\lambda_t|^2+\frac{4}{3}c{}_1^b|\lambda_b|^2-\frac{4}{3}c{}_1^{b^c}|\lambda_{b^c}|^2+4c{}_1^{\rm A}g_3^2} \nonumber\\
m_S^2 & \approx\frac{g_\rho^2}{16\pi^2}f^2\nbrack{-\frac{14}{3}c{}_1^\chi|\lambda_\chi|^2+4c{}_1^b|\lambda_b|^2}
\end{align}
generically predicting a triplet mass
\be
m_T\sim\frac{g_\rho}{4\pi}\max\sbrack{|\lambda_\psi|,g_3}f
\ee 
where $\psi = \chi, t, b,b^c$.  When $|\lambda_\chi|\lesssim|\lambda_b|$ the singlet mass is approximated by
\be
m_S\sim\frac{g_\rho}{4\pi}|\lambda_b|f\sim\frac{g_\rho}{4\pi}\frac{|\lambda_b|}{|\lambda_\chi|}m_\chi
\ee
and the singlet is heavier than or similar in mass to the top companions, which have mass $m_\chi\sim|\lambda_\chi|f$.  When $|\lambda_\chi|\gtrsim|\lambda_b|$ the singlet mass is approximated by
\be
m_S\sim\frac{g_\rho}{4\pi}|\lambda_\chi|f\sim\frac{g_\rho}{4\pi}m_\chi
\ee
and the singlet is lighter than the top companions (in all of the above expressions we have dropped all order-one coefficients, including the $c$'s).  When $|\lambda_\chi|\sim|\lambda_b|$ it is also possible to have a cancellation in the expression for the singlet mass, i.e.\ a second tuning, to give
\be
m_S\ll\frac{g_\rho}{4\pi}|\lambda_\chi|f\lesssim m_\chi
\ee
whereupon the singlet is much lighter than the top companions.

\section{Dark matter\label{sec:DM}}

\subsection{Dark matter stability}

As it stands there are four global $U(1)$ symmetries in the model left unbroken by the strong sector.  They are $U(1)_7$ and $U(1)_6$, both subgroups of $SU(7)$, and $U(1)_B$ and $U(1)_L$.  Coupling the left-handed quark doublet to two different strong sector operators breaks one linear combination, coupling the left-handed lepton doublet to two different strong sector operators breaks a second linear combination, the Higgs VEV breaks a third and the Majorana mass for the right-handed neutrino breaks a fourth.  The charges of all states under the $U(1)$ symmetries are given in table~\ref{tab:dmcharges} and it is easy to check that four independent symmetries are broken, leaving no symmetries to stabilise our scalar singlet, dark matter candidate.

\begin{table}[!t]
\be\nonumber
\begin{array}{|l|rrrr|rr|}\hline
& U(1)_{\slashed{q}} & U(1)_{\slashed{l}} & U(1)_{\slashed{H}} & U(1)_L & U(1)_B & Z_3 \\\hline
T & 0 & 0 & -2 & 0 & 0 & -1 \\
D & 0 & 0 & -2 & 0 & 0 & 0 \\
S & 0 & 7 & 10 & 0 & \frac{1}{3} & 1 \\
q_{(u)} & -1 & 6 & 11 & 0 & \frac{1}{3} & 0 \\
q_{(d)} & 1 & 6 & 11 & 0 & \frac{1}{3} & 0 \\
u^c & 1 & -6 & -9 & 0 & -\frac{1}{3} & 0 \\
d^c & -1 & -6 & -13 & 0 & -\frac{1}{3} & 0 \\
l_{(\nu)} & 0 & 0 & 2 & 1 & 0 & 0 \\
l_{(e)} & 0 & 2 & 2 & 1 & 0 & 0 \\
N^c & 0 & 0 & 0 & -1 & 0 & 0 \\
e^c & 0 & -2 & -4 & -1 & 0 & 0 \\
\tilde{q}^c & -1 & 6 & 9 & 0 & \frac{1}{3} & -1 \\
\tilde{e} & -1 & 6 & 9 & 0 & \frac{1}{3} & 1 \\
\tilde{d}^c & -1 & -1 & -3 & 0 & 0 & 1 \\
\tilde{l} & -1 & -1 & -3 & 0 & 0 & 0 \\\hline\end{array}
\ee
\caption{Charges of all states under the global symmetries preserved by the strong sector, derived from eq.~\eqref{eq:U1defs} and table~\ref{tab:SU7}.  Baryon number and the associated baryon triality, $Z_3\equiv3U(1)_B-n_c\mbox{ mod 3}$, are also preserved by the elementary sector (modulo sphaleron processes and gravitational effects).\label{tab:dmcharges}}
\end{table}

To generate one we will enlarge the global symmetry in the strong sector from $SU(7)$ to $U(7)\equiv SU(7)\times U(1)_E$.  $U(1)_E$ acts as an external symmetry to $SU(7)$, with the charge of each representation given by the number of fundamental indices it carries minus the number of antifundamental.  It is completely analogous to baryon number in QCD\@.  In addition we will allow the true baryon number to be a mixture of an external baryon number, $U(1)_{B_0}$, and the other $U(1)$ symmetries respected by the strong sector.  It is then convenient to work in the basis
\begin{align}
U(1)_{\slashed{q}} & \equiv\frac{1}{3}\sbrack{U(1)_E+2U(1)_L} \nonumber\\
U(1)_{\slashed{l}} & \equiv\frac{1}{6}\sbrack{4U(1)_E+U(1)_7-7U(1)_6+90U(1)_{B_0}+8U(1)_L} \nonumber\\
U(1)_{\slashed{H}} & \equiv\frac{1}{3}\sbrack{5U(1)_E-6U(1)_6+90U(1)_{B_0}+10U(1)_L} \nonumber\\
U(1)_B & \equiv\frac{1}{126}\sbrack{6U(1)_E+U(1)_7-7U(1)_6+126U(1)_{B_0}}. \label{eq:U1defs}
\end{align}
Lepton number is defined as usual and $U(1)_{B_0}$ is the baryon number symmetry external to $U(7)$, under which $q_{(u)}$, $l_{(\nu)}$, $(\tilde{q}^c,\tilde{e})$ and $(\tilde{d}^c,\tilde{l})$ are assigned charge $\frac{1}{3}$, and $u^c$ and $N^c$ are assigned charge $-\frac{1}{3}$.  All other states are neutral.  $U(1)_{B_0}$ mixes with symmetries internal to $U(7)$ to give the true baryon number symmetry, $U(1)_B$, defined above.

This basis is convenient as it is immediately apparent from table~\ref{tab:dmcharges} that $U(1)_{\slashed{q}}$ is broken only by coupling the left-handed quark doublet to two different strong sector operators, i.e.\ the left-handed quark, $q$, has two different charges under $U(1)_{\slashed{q}}$ but only one charge under all other $U(1)$ symmetries.  Similarly, $U(1)_{\slashed{l}}$ is broken only by coupling the left-handed lepton doublet to two different strong sector operators and $U(1)_{\slashed{H}}$ is broken only by the Higgs VEV\@.  Each of these symmetries is broken to a $Z_2$ subgroup.  True baryon number remains unbroken so there are no issues with proton stability.  It is also apparent that an unbroken, baryon triality symmetry
\be
Z_3\equiv3U(1)_B-n_c\mod{3}
\ee
exists in this model, under which no SM states carry charge.

By allowing baryon number to mix with the $U(1)$ symmetries internal to the strong sector's global $SU(7)$ symmetry we find that the scalar singlet, $S$, has picked up a non-zero baryon number and, consequently, is charged under baryon triality.  As long as the scalar singlet is lighter than the scalar triplet and the top companions, baryon triality therefore renders it stable and it acts as pNGB dark matter~\cite{Frigerio:2012uc}.

\subsection{Experimental limits}

Based on the expressions \eqref{eq:TSmasses} for the pNGB masses and recalling that $m_\chi\sim|\lambda_\chi|f$ there are two situations in which the singlet can be lighter than the triplet and top companions.  Either $|\lambda_\chi|\gtrsim|\lambda_b|$ or $|\lambda_\chi|\sim|\lambda_b|$ and there is an additional tuning in the expression for the mass of the scalar singlet that results in $m_S\ll(g_\rho/4\pi)m_\chi$.

\begin{figure}[!t]
\begin{center}
\includegraphics[height=3.1cm]{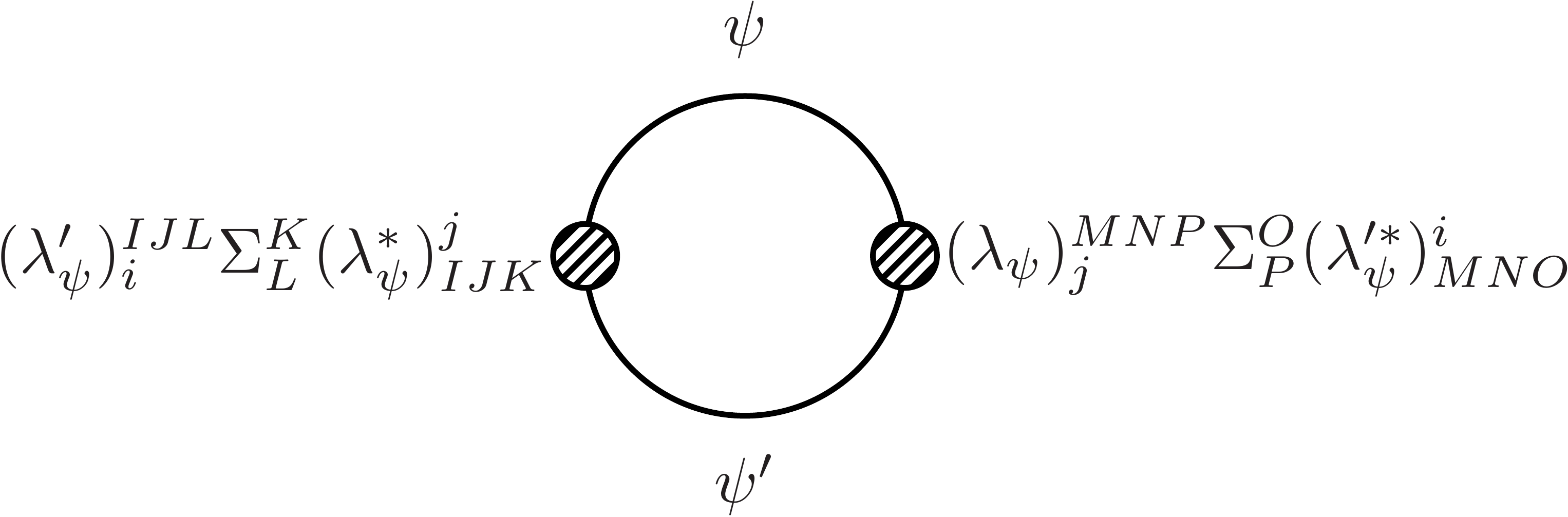}
\caption{Elementary fermion loops that generate next-to-leading order contributions to the pNGB potential.\label{fig:lambda4}}
\end{center}
\end{figure}

In the limit of large $f$ (specifically $f\gg m_S$) we can ignore all other couplings of the scalar singlet when calculating the relic density and keep only the Higgs-portal coupling, $V\supset\kappa|D|^2|S|^2$ \cite{Frigerio:2012uc}.  This arises at one-loop order in the elementary fermions and at quartic order in the projectors.  It is generated by diagrams like the one shown in figure~\ref{fig:lambda4} and is given by
\be
\kappa=\frac{1}{16\pi^2}\nbrack{\frac{28}{9}c{}_2^{\chi\chi}|\lambda_\chi|^4+\frac{4}{3}c{}_2^{bb}|\lambda_b|^4-\frac{4}{3}c{}_2^{bb^c}|\lambda_b|^2|\lambda_{b^c}|^2+\frac{2}{3}c{}_2^{tb}|\lambda_t|^2|\lambda_b|^2}
\ee
the full derivation being given in appendix~\ref{app:VpNGB}.

In ref.~\cite{Cline:2013gha} it was shown that the relic density of a complex scalar singlet, coupling only via the Higgs-portal does not overclose the universe provided
\be\label{eq:kappacon}
\kappa\gtrsim3\times10^{-4}\pfrac{m_S}{\rm GeV}.
\ee
Taking $|\lambda_\chi|\gtrsim|\lambda_b|$, so that the scalar singlet can be lighter than the top companions, using $|\lambda_b||\lambda_{b^c}|\sim y_bg_\rho$, and substituting in the natural value for the scalar singlet mass in this scenario, $m_S\sim g_\rho|\lambda_\chi|f/(4\pi)$, the constraint becomes
\be
\max\sbrack{|\lambda_\chi|^4,y_b^2g_\rho^2}\gtrsim g_\rho|\lambda_\chi|\pfrac{f}{\rm TeV}.
\ee
There are no viable solutions when $|\lambda_\chi|^4<y_b^2g_\rho^2$ (assuming that $|\lambda_b|\sim|\lambda_{b^c}|$) as the constraint can only be satisfied if
\be
g_\rho\gtrsim\frac{1}{y_b^3}\pfrac{f}{\rm TeV}^2\gg4\pi
\ee
thereby violating the upper bound on $g_\rho$.  When $|\lambda_\chi|^4>y_b^2g_\rho^2$ the constraint can be satisfied if
\be
\frac{|\lambda_\chi|}{g_\rho}\gtrsim\pfrac{f}{g_\rho^2\mbox{ TeV}}^{\frac{1}{3}}\gtrsim\pfrac{f}{16\pi^2\mbox{ TeV}}^{\frac{1}{3}}
\ee
but this strains the assumption that $\lambda_\chi$ can be treated as a perturbation to the strong sector for larger values of $f$.  However, it should be noted that $m_S$ is starting to become comparable to $f$ for such large values of $|\lambda_\chi|$ so other annihilation channels will start to turn on, reducing the relic density.

Alternatively we can allow for an additional tuning that makes the scalar singlet light.  We have already allowed one tuning in the model, which is physically motivated by a desire to keep the Higgs VEV low.  The upper limit we must impose on $m_S$, so that the scalar singlet does not overclose the universe, provides a second, physical motivation for tuning.  Whatever argument is ultimately invoked to explain the tuning in the Higgs sector, is therefore equally applicable here.

The tuning in question is accomplished by choosing $|\lambda_\chi|\sim|\lambda_b|$, then tuning the $c$'s to arrange for a cancellation in the expression given in eq.~\eqref{eq:TSmasses}.  The mass of the scalar singlet may now be treated as a free parameter.  When $|\lambda_\chi|^4<y_b^2g_\rho^2$ there are still no viable solutions to eq.~\eqref{eq:kappacon} as $m_S$ can be no more than a GeV and the model is ruled out by direct detection experiments \cite{Cline:2013gha}.  We therefore take $|\lambda_\chi|^4>y_b^2g_\rho^2$, whereupon the Higgs-portal coupling is given by
\be
\kappa\sim0.02\pfrac{m_\chi}{f}^4
\ee 
upon replacing $\lambda_\chi$ with $m_\chi\sim\lambda_\chi f$.

\begin{figure}[!t]
\begin{center}
\includegraphics[width=0.45\textwidth]{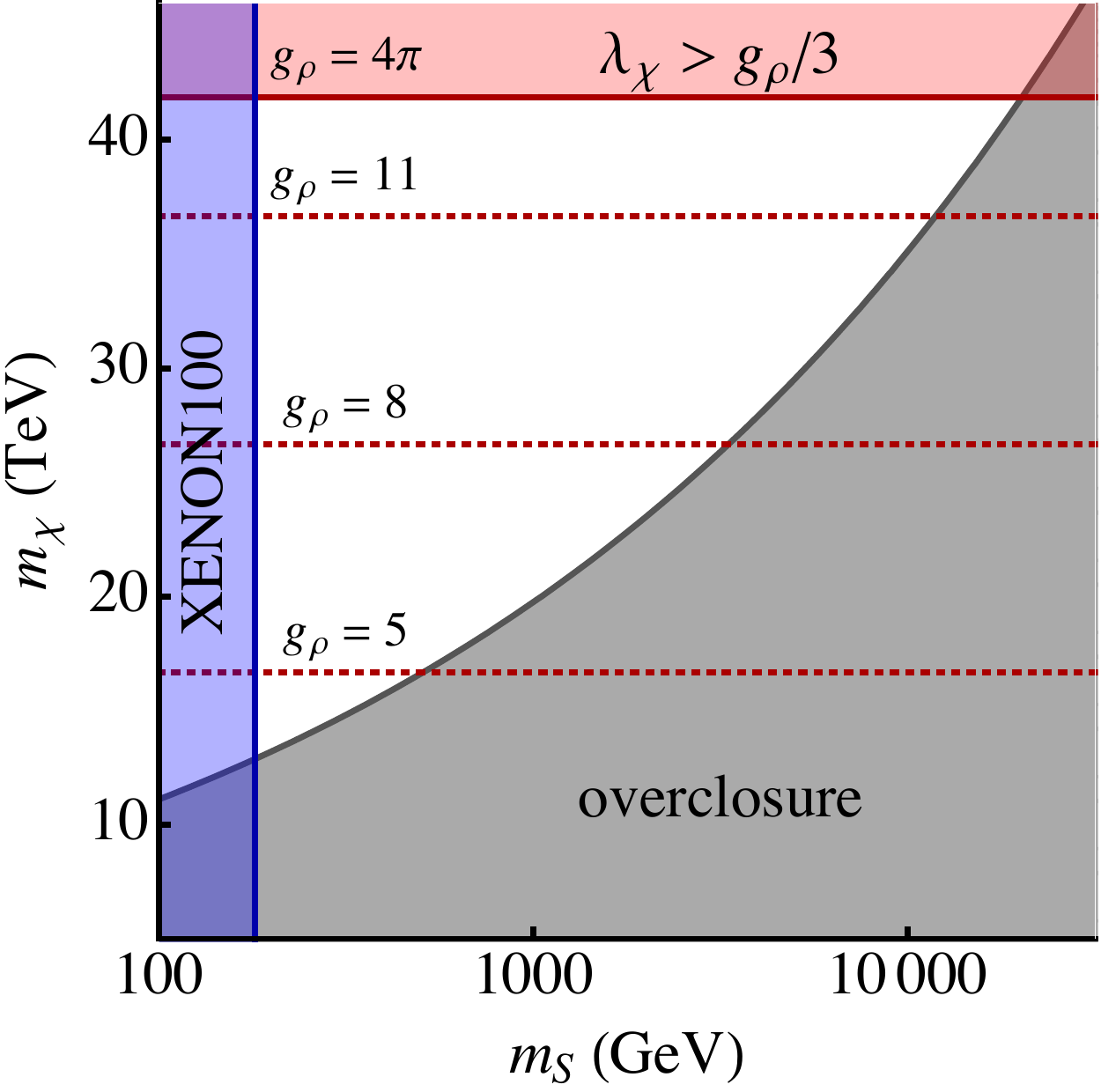}
\caption{The allowed parameter space in the $m_S-m_\chi$ plane with $f=10$~TeV\@.  In the lower region the scalar singlet overcloses the universe.  In the left-hand region the model is excluded by direct detection experiments.  This region will expand to the right as experiments become more sensitive.  In the upper regions $\lambda_\chi/g_\rho$ is not small enough for our calculations to be trusted (contours for various values of $g_\rho$ are plotted).\label{fig:mSmx}}
\end{center}
\end{figure}

Using the results given in ref.~\cite{Cline:2013gha} to investigate which values of $m_S$ and $m_\chi$ are consistent with observations leads to the results summarised in figure~\ref{fig:mSmx}.  We have assumed that $2m_S>m_h$ such that resonant annihilation does not occur.  Eq.~\eqref{eq:kappacon} imposes an upper limit on the scalar singlet mass for a given value of top companion mass; it is difficult to raise the dark matter mass above around 10~TeV without overclosing the universe.  Direct detection experiments place a lower bound on the scalar singlet mass of around 180~GeV, which will increase as experiments become more sensitive~\cite{Aprile:2012nq}.  While not necessarily a physical limit our calculations become unreliable when $|\lambda_\chi|\gtrsim g_\rho/3$, so we will discount models with heavier top companions.  Combining all the constraints we find that a realistic model with $f=10$~TeV should have
\begin{align}
180\mbox{ GeV}\lesssim m_S\lesssim10\mbox{ TeV} && 10\mbox{ TeV}\lesssim m_\chi\lesssim40\mbox{ TeV.}
\end{align}
Having applied all constraints, the fact that an allowed range for the top companion mass still exists is both non-trivial and encouraging.  Scaling $f$ simply scales this range linearly; the preferred range for the scalar singlet mass remains unchanged.

The tuning in the scalar singlet mass is quantified by
\be
\frac{m_S}{g_\rho|\lambda_\chi|f/(4\pi)}\lesssim0.3\pfrac{m_\chi}{\rm TeV}^2\pfrac{\rm TeV}{f}^3
\ee
where the upper bound comes from imposing eq.~\eqref{eq:kappacon} and $|\lambda_\chi|\lesssim g_\rho/3$.  For $f=10$~TeV the least tuned scenario is when $m_S\approx10$~TeV and $m_\chi\approx40$~TeV, and the tuning is only around 25\%, requiring $g_\rho=4\pi$.  The most tuned scenario is when $m_S=180$~GeV and $m_\chi\approx10$~TeV, and the tuning is around 2 or 3\% for $g_\rho=4\pi$ and $g_\rho=8$ respectively.

\section{Exotic state phenomenology\label{sec:ES}}

Several exotic states, namely the scalar triplet, $T$, and the top companions, $\tilde{q}^c$, $\tilde{e}$, $\tilde{d}^c$ and $\tilde{l}$, have been introduced in this model.  Their SM charges under $SU(3)\times SU(2)\times U(1)_Y$ are
\begin{align}
T & \in({\bf3},{\bf1})_{-\frac{1}{3}} &
\tilde{q}^c & \in(\overline{\bf3},{\bf2})_{-\frac{1}{6}} &
\tilde{e} & \in ({\bf1},{\bf1})_{-1} &
\tilde{d}^c & \in (\overline{\bf3},{\bf1})_{\frac{1}{3}} &
\tilde{l} & \in ({\bf1},{\bf2})_{-\frac{1}{2}}.
\end{align}
In this section we will investigate how these states decay.  The decays must be generated by the strong sector and the large global symmetry group will generally restrict them to proceed through operators with large dimensions.  This raises the possibility of states that are long-lived or collider-stable at the LHC and future high-energy experiments.  For concreteness, we choose $f=10$~TeV then assume the spectrum
\begin{align}
m_\chi & \sim\text{(1--2)}f \sim\text{10--20 TeV} & m_T & \sim\text{(1--2)}\frac{f}{\pi}\sim\text{3--5 TeV} & m_S & \lesssim\text{1 TeV}
\end{align}
which is consistent with the analysis above.  Any other spectrum with $f$ in the 10--1000~TeV range would be equally valid, although the prospect of discovery in future experiments diminishes with increasing $f$.  In taking the top companions, $\chi$ to be roughly degenerate we will consider only their decays to $T$, $S$ and SM particles.

All of the coloured top companions have unsuppressed decays.  The $SU(2)$ doublet has a two-body decay at quadratic order in the $\lambda$'s through the coupling
\be
\cL\supset \frac{1}{3f} \, |\lambda{}_\chi^{\bf10}||\lambda_t| \, \Pi^{\chi t} \, (T\bar{\tilde{q}}^c\slashed{p}q)
\ee
(here and below brackets denote gauge index contraction).  This is generated by the term
\be
\cL\supset \Pi^{\chi t}\sbrack{(\lambda{}_\chi^{{\bf10}*})^{i_4i_2}_{IJK}(\lambda_t)^{j_3j_2,IJL}\Sigma{}^K_L}(\bar{\tilde{q}}^c,\bar{\tilde{e}})_{i_4i_2}\slashed{p}q_{j_3j_2}
\ee
from eq.~\eqref{eq:Lefff}.  The $SU(2)$ singlet has a three-body decay at linear order in the $\lambda$'s via
\be
\cL\supset  \frac{1}{3f} \, |\lambda{}_\chi^{\bf5}| \, S \, (T^\dag\tilde{d}^ct^c)
\ee
generated by the term
\be
\cL\supset(\lambda{}_\chi^{\bf5})^{IJL}_{i_5}\Sigma{}^K_L(\tilde{d}^c,\tilde{l})^{i_5}(\cO_t)_{IJK},
\ee
from eq.~\eqref{eq:Lmix}, where $(\cO_t)_{IJK} = f (t^c)^{i_3} (\lambda_{t^c})_{i_3 IJK}$.  Both decays are prompt on collider timescales.

The uncoloured top companions do not decay at quadratic level in the $\lambda$'s.  For the electroweak singlet the leading decay is to a bottom quark and two scalar triplets through the next-to-leading order coupling
\be\label{eq:etdecay}
\cL\supset\frac{c{}_4^{\tilde{e}}}{18(4\pi)^2m_\rho} \, |\lambda{}_\chi^{\bf10}||\lambda_t||\lambda_b||\lambda_{b^c}| \, \tilde{e}(b^cT^\dag T^\dag)
\ee
generated by
\be
\cL\supset \frac{1}{f^3} \, \Pi{}_4^{\tilde{e}}\sbrack{(\lambda{}_\chi^{\bf10})_{i_4i_2}^{IJL}(\lambda_t^*)_{i_3j_2,IKM}(\lambda_b)^{j_3k_2}_{JNO}(\lambda_{b^c})_{l_3}^{MOP}\Sigma{}_L^K\Sigma{}_P^N}(\tilde{q}^c,\tilde{e})^{i_4i_2}\bar{q}^{i_3j_2}\slashed{p}q_{j_3k_2}b^c{}^{l_3}
\label{eq:etoperator}\ee
after closing off the quark loop.\footnote{It is not possible to generate the operator \eqref{eq:etdecay} at quadratic order in the $\lambda$'s, as it requires two factors of the spurion, $\Sigma$.}  $\Pi{}_4^{\tilde{e}}$ is a strong sector form factor and $c{}_4^{\tilde{e}}$ an order-one coefficient coming from the loop integral
\be
\int\frac{\text{d}^4p}{(2\pi)^4}\, \Pi{}_4^{\tilde{e}} (p^2) = \frac{c{}_4^{\tilde{e}}}{g_\rho} \frac{f^4}{16\pi^2}.
\label{eq:eint}\ee

\begin{figure}
\centering
\includegraphics[width=0.4\textwidth]{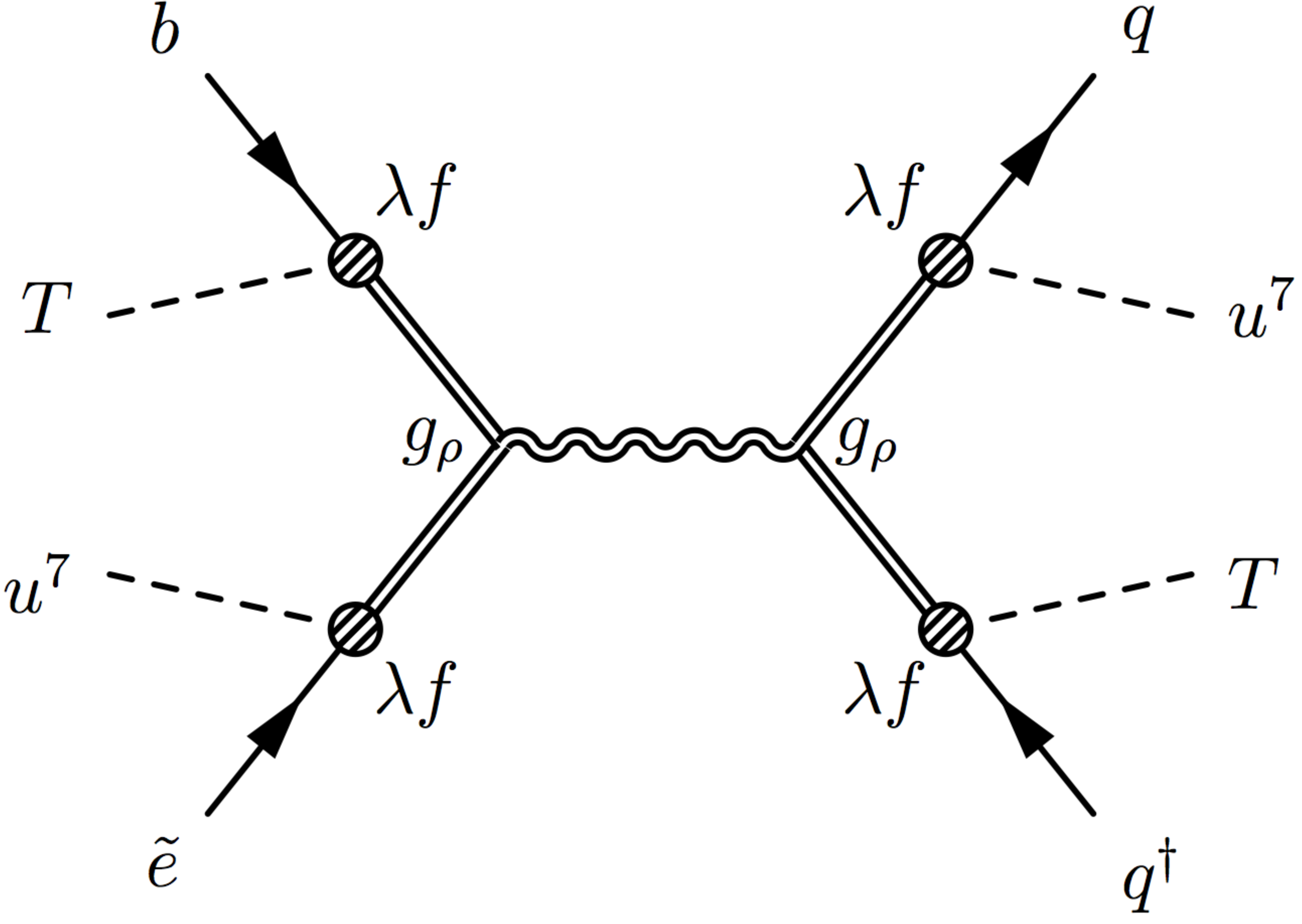}
\caption{Diagram contributing to the generation of eq.~\eqref{eq:etoperator}.  Double lines represent composite sector resonances.  Integrating out this diagram gives a parametric contribution $\sim g_\rho^2 f^4 m_\rho^{-7} \sim 1/(g_\rho^5f^{3})$.}\label{fig:fourspurion}
\end{figure}

The parametric dependence is understood by considering diagrams like figure~\ref{fig:fourspurion}, which are responsible for generating eq.~\eqref{eq:etoperator}.  This diagram has four composite fermion propagators, one boson propagator and four factors of $f$ from the elementary-composite mixing.  Since we need to extract a factor of $\slashed{p}$ to contract the spinors in eq.~\eqref{eq:etdecay} the diagram scales as
\begin{equation}
  g_\rho^2 f^4 \, \frac{1}{m_\rho^2} \, \pfrac{1}{m_\rho}^3 \, \frac{\slashed{p}}{m_\rho^2} \sim \frac{\slashed{p}}{g_\rho^5f^3} \,.
\end{equation}
at low energy.  It follows that the form factor $\Pi{}_4^{\tilde{e}}$ scales like $g_\rho^{-5}$.  The resultant lifetime is then
\begin{align}
  \Gamma (\tilde{e} \to \bar{b} T T) & \approx \frac{(c{}_4^{\tilde{e}})^2}{2^{5} 3^4 (4\pi)^7}\pfrac{|\lambda{}_\chi^{\bf10}||\lambda_t||\lambda_b||\lambda_{b^c}|}{g_\rho}^2 \frac{m_\chi^3}{f^2} \notag \\
  c \tau & \sim 10^{-7} \, \text{cm} \, \pfrac{1}{\smash{c{}_4^{\tilde{e}}}}^2 \pfrac{\text{20 TeV}}{m_\chi}^5 \pfrac{f}{\text{10 TeV}}^4,
\end{align}
using $|\lambda_t| \approx 3 y_t$ and $|\lambda_b||\lambda_{b^c}| \approx 3 g_\rho y_b$.  Decays are collider-prompt unless $m_\chi$ is very small; either $m_T \ll m_\chi \lesssim 3$~TeV or $m_\chi \sim m_T$ so that there is an additional phase-space suppression.  The operator in eq.~\eqref{eq:etoperator} will also lead to a five-body decay mode but this is suppressed by both phase space and SM quark masses so the loop decay mode dominates.

Examining eq.~\eqref{eq:etoperator} more carefully we note that this operator is not actually unique.  There are multiple possible contractions of the fermion indices, plus the momentum $\slashed{p}$ can be any combination of the external momenta.  It follows that the integrand in eq.~\eqref{eq:eint} will also contain additional (dimensionless) kinematic factors from the fermion loop, and that $c{}_4^{\tilde{e}}$ is an order-one form factor rather than a constant.  Without a calculable theory of the strong sector we can do no better than the order-one estimates used here.  However, the main result, that $\tilde{e}$ decays promptly, is robust to these approximations.

The $SU(2)$ doublet has similar phenomenology.  It decays to a quark, a scalar triplet and a scalar singlet through the next-to-leading order coupling
\be
\cL\supset\frac{c{}_4^{\tilde{l}}}{16\pi^2m_\rho} \, |\lambda{}_\chi^{\bf5}||\lambda_b||\lambda_\nu||\lambda_\tau| \, S^\dag(\tilde{l}qT^\dag)
\label{eq:ltdecay}\ee
generated by
\begin{align}
  \cL \supset \Pi{}_4^{\tilde{l}} \bigl\{ & \sbrack{(\lambda{}_\chi^{\bf5})_{i_5}^{IJL}(\lambda_b)^{i_3i_2}_{IJK}\Sigma{}_L^K} \sbrack{(\lambda_\nu^*)_{j_2,MN}(\lambda_\tau)^{k_2,MO}\Sigma{}_O^N} \notag \\
  +& \sbrack{(\lambda{}_\chi^{\bf5})_{i_5}^{IJL}(\lambda_b)^{i_3i_2}_{IKM}\Sigma{}_L^K \; (\lambda_\nu^*)_{j_2,JN}(\lambda_\tau)^{k_2,MO}\Sigma{}_O^N} \bigr\} \, (\tilde{d}^c,\tilde{l})^{i_5}q_{i_3i_2} \; \bar{l}^{j_2}\slashed{p}l_{k_2}
\end{align}
after closing off the lepton loop.  $\Pi{}_4^{\tilde{l}}$ is a strong sector form factor, different to $\Pi{}_4^{\tilde{e}}$ but with the same parametric dependence on $g_\rho$.  The loop decay is again much quicker than the five-body decay and $c{}_4^{\tilde{l}}$ is an order-one form factor defined analogously to $c{}_4^{\tilde{e}}$.  This decay involves a lepton loop, leading to a suppression by lepton Yukawa couplings, but this is compensated by larger group-theoretical factors in eq.~\eqref{eq:ltdecay} as compared to eq.~\eqref{eq:etdecay}.  We find
\begin{align}
  \Gamma (\tilde{l} \to \bar{q} T S^\dagger) & \approx \frac{(c_4^{\tilde{l}})^2}{2^3 (4\pi)^7} \pfrac{|\lambda{}_\chi^{\bf5}||\lambda_b||\lambda_\nu||\lambda_\tau|}{g_\rho}^2 \frac{m_\chi^3}{f^2} \notag \\
  c \tau & \sim 10^{-7} \, \text{cm} \, \pfrac{1}{\smash{c{}_4^{\tilde{l}}}}^2 \pfrac{8}{g_\rho}\pfrac{\text{20 TeV}}{m_\chi}^5 \pfrac{f}{\text{10 TeV}}^4
\end{align}
where we have assumed that $\lambda_\nu \sim \lambda_\tau \sim \sqrt{2g_\rho y_\tau}$ and $\lambda_b \sim \sqrt{3g_\rho y_b}$.  As before this is collider-prompt unless $m_T \ll m_\chi\lesssim 3$~TeV or $m_\chi \sim m_T$.

\begin{figure}
\centering
\includegraphics[width=0.4\textwidth]{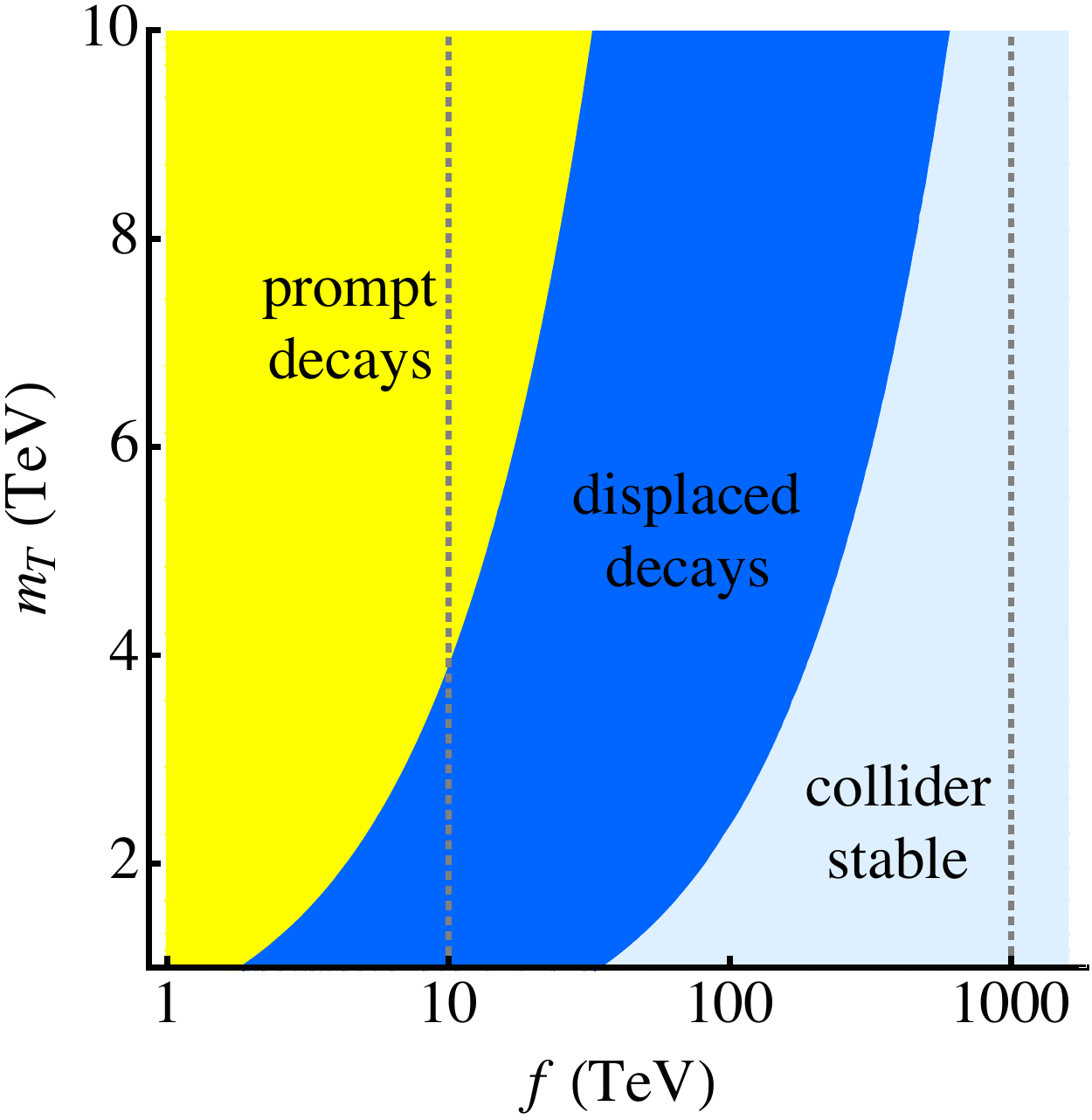}
\caption{The scalar triplet decay phenomenology as a function of $f$ and $m_T$.  Displaced decays correspond to lifetimes in the range 100~$\mu$m$\,\leq c\tau \leq 10$~m.  The dotted lines bound the range of $f$ favoured by precision-electroweak and flavour constraints, $f\gtrsim10$~TeV, and gauge coupling unification, $f\lesssim100$--1000~TeV.}
\label{fig:ctauT}
\end{figure}

The scalar triplet does not decay at quadratic order in the projectors either.  This can be understood through the residual $Z_2$ symmetry that remains when $U(1)_{\slashed{l}}$ is broken, as discussed in section~\ref{sec:DM}.  The combination of this symmetry and baryon triality forces the scalar triplet to decay to two scalar singlets, again requiring two insertions of $\Sigma$.  The leading decay mode is to a top quark, a bottom quark and two scalar singlets through the coupling
\be
\cL\supset \frac{c{}_3^T}{24\pi^2f^2} \, |\lambda_{b^c}||\lambda_\nu||\lambda_\tau|\, S^2(T^\dag t^cb^c) \,.
\ee
This is generated at next-to-leading order in the projectors by the coupling
\be
\cL\supset \frac{1}{f^4} \, \Pi{}_3^T \sbrack{(\lambda_{b^c})^{IJL}_{i_3}\Sigma{}^K_L}b^c{}^{i_3}(\cO_t)_{IJK}\sbrack{(\lambda_\nu)^{MO}_{i_2}(\lambda_\tau^*)_{MN}^{j_2}\Sigma{}^N_O}\bar{l}^{i_2}\slashed{p}l_{j_2} 
\ee
after closing off the lepton loop.  The form factor has the dependence $\Pi{}_3^T \sim g_\rho^{-4}$, which is different to the $\Pi_4$'s because a spurion is replaced with the composite sector operator $\cO_t$.  Assuming that we can neglect the scalar singlet mass, the lifetime is
\begin{align}
  \Gamma (T \to \bar{t} \bar{b} S S) & \approx \frac{(c{}_3^T)^2}{2^{6} 3^4 5 (2\pi)^{9}} \nbrack{|\lambda_{b^c}||\lambda_\nu||\lambda_\tau|}^2 \frac{m_T^5}{f^4} \notag \\
  c \tau & \approx 0.2 \, \text{mm} \, \pfrac{1}{\smash{c{}_3^{T}}}^2 \pfrac{8}{g_\rho}^3 \pfrac{\text{3 TeV}}{m_T}^5 \pfrac{f}{\text{10 TeV}}^4 \,.
\end{align}
This can lead to decays that are prompt, displaced or collider stable.  For our canonical values of $m_T$ and $f$ the decays are either prompt or displaced.  The regions of parameter space that lead to different phenomenology are shown in figure~\ref{fig:ctauT}.

Since the scalar triplet is the lightest, coloured exotic state predicted by our model it will generally be the most promising state to search for at colliders.  It will be pair produced (owing to the baryon triality symmetry) via pure-QCD processes and has the same quantum numbers as a scalar bottom quark, so searches designed for supersymmetric models can also be of use here.  When the scalar triplet is long-lived, R-hadron searches \cite{Aad:2012pra, Chatrchyan:2013oca} can be applied and the current limit on the scalar triplet mass is between 800~GeV and 1~TeV\@.  When the scalar triplet decays promptly the collider signature is two top quarks, two bottom quarks and missing energy from the scalar singlets.  This signature is covered by the gluino search in ref.~\cite{Aad:2014lra}, which imposes limits of around 1.3~TeV on the gluino mass.  We expect similar limits to apply to our model, as long as the scalar singlet is lighter than around 700 GeV\@.

\section{Conclusion\label{sec:CO}}

By simply allowing for a large scale of spontaneous symmetry breaking, $f\gtrsim 10$ TeV, all precision-electroweak and flavour constraints on composite Higgs models can be trivially satisfied.  This requires a tuning of order $10^{-4}$ in the Higgs potential to obtain the observed Higgs boson mass, but no additional symmetries, such as a custodial symmetry or flavour symmetries, are needed in the strong sector.  The tuning produces a `split' spectrum of composite states, where the resonance masses are in the 10--1000~TeV range and the pNGBs remain near the electroweak scale.

Even though the composite Higgs model is unnatural, the strong dynamics underlying the compositeness helps address other shortcomings of the Standard Model.  The fermion mass hierarchy is explained by assuming that the right-handed top quark is fully composite and the remaining SM fermions are mostly elementary.  This idea of partial compositeness also introduces new contributions to the running of the SM gauge couplings and, intriguingly, improves unification when compared to the SM alone.  The new contributions, due to fermionic top companions, depend on the scale $f$, so arbitrarily raising the scale worsens the unification.  Restricting the corrections required for gauge coupling unification to be compatible with precision unification then leads to an upper bound on the symmetry breaking scale, $f\lesssim100$--1000~TeV, the range due to the uncertainty in estimating higher-loop contributions.  This implies that strong sector resonances may be seen at future colliders or lead to effects in rare decay experiments.

A dark matter candidate can be provided by the strong sector and identified with one of the pNGBs.  The minimal coset space containing an unbroken $SU(5)$ symmetry and a stable singlet to act as dark matter is $SU(7)/SU(6)\times U(1)$.  The resulting model is compatible with direct detection experiments provided the scalar singlet mass is greater than 180 GeV and the top companion masses are several tens of TeV\@.  Furthermore, the colour-triplet partner of the Higgs in the GUT multiplet, is long-lived due to the large global symmetry group.  At leading order it decays via a dimension-six operator.  This can lead to a displaced vertex when produced at a collider, either the LHC or a future collider, providing a distinctive experimental signature.

The global symmetries of the strong sector also include a baryon number symmetry to prevent rapid proton decay due to composite states.  The only potential source of proton decay is then from the elementary sector and is suppressed by at least the GUT scale, $10^{15}$ GeV, and will almost always come with a much larger suppression.  Similarly, lepton number is preserved by the strong sector so there are no large contributions to neutrino masses.  However, lepton number must be broken by the introduction of a right-handed neutrino in the elementary sector to ensure that the scalar singlet is the only stable, dark matter candidate.  The SM neutrino masses are then explained by a type-I see-saw mechanism and the matter-antimatter asymmetry of the universe can be generated via leptogenesis in the elementary sector.

Clearly the shortcomings of the SM can be straightforwardly addressed with an unnatural composite Higgs model, provided there is a modest amount of tuning in the Higgs potential.  This is still a many-orders-of-magnitude improvement on the usual SM tuning, with the added benefit that the strong dynamics provides new physics to explain the fermion mass hierarchy, dark matter and the unification of the gauge couplings.  Even though the origin of the tuning remains obscure, and yet to be understood from the underlying strong dynamics, the low energy predictions can be tested at the LHC and future experiments.  It may very well provide a model of our universe or, at the very least, some other place in the multiverse.

\section*{Acknowledgements}

We thank Alex Pomarol for helpful discussions.  This work was supported by the Australian Research Council.  TG was also supported by the Department of Energy grant DE-SC0011842. TG thanks the hospitality of the Aspen Center for Physics and the partial support from National Science Foundation Grant No\@. PHYS-1066293 during the completion of this work.  JB is grateful to the Mainz Institute for Theoretical Physics (MITP) for its hospitality and partial support during the completion of this work.  JB, TSR and AS thank the Fine Theoretical Physics Institute for partial support and hospitality during the completion of this work. TSR acknowledges the INSPIRE Faculty Grant, Department of Science and Technology, India.

\appendix
\newpage

\numberwithin{table}{section}

\section{$U(7)$ representations\label{app:U7}}

Several representations of $U(7)\equiv SU(7)\times U(1)_E$ are made use of in the text.  These decompose into $SU(6)\times U(1)_E\times U(1)_7$ and $SU(5)\times U(1)_E\times U(1)_7\times U(1)_6$ representations as detailed in table~\ref{tab:SU7}.  $U(1)_E$ is the $U(1)$ symmetry contained within $U(7)$ but not $SU(7)$, $U(1)_7$ is the $U(1)$ symmetry contained within $SU(7)$ but not $SU(6)$ and $U(1)_6$ is the $U(1)$ symmetry contained within $SU(6)$ but not $SU(5)$.  All of these properties and more can be conveniently derived using {\tt LieART} \cite{2012arXiv1206.6379F}.

\begin{table}[!b]
\be\nonumber
\begin{array}{|l|lrr|lrrr|}\hline
U(7) & SU(6) & U(1)_E & U(1)_7 & SU(5) & U(1)_E & U(1)_7 & U(1)_6 \\\hline\hline
{\bf7}=\fund & {\bf6}=\fund & 1 & 1 & {\bf5}=\fund & 1 & 1 & 1 \\
&&&& {\bf1} & 1 & 1 & -5 \\
& {\bf1} & 1 & -6 & {\bf1} & 1 & -6 & 0 \\\hline
{\bf21}=\asym & {\bf15}=\asym & 2 & 2 & {\bf10}=\asym & 2 & 2 & 2 \\
&&&& {\bf5}=\fund & 2 & 2 & -4 \\
& {\bf6}=\fund & 2 & -5 & {\bf5}=\fund & 2 & -5 & 1 \\
&&&& {\bf1} & 2 & -5 & -5 \\\hline
{\bf35}=\asymT & {\bf20}=\asymT & 3 & 3 & {\bf10}=\asym & 3 & 3 & -3 \\
&&&& \overline{\bf10}=\overline{\asym} & 3 & 3 & 3 \\
& {\bf15}=\asym & 3 & -4 & {\bf10}=\asym & 3 & -4 & 2 \\
&&&& {\bf5}=\fund & 3 & -4 & -4 \\\hline
{\bf48}={\rm adj} & {\bf35}={\rm adj} & 0 & 0 & {\bf24}={\rm adj} & 0 & 0 & 0 \\
&&&& {\bf5}=\fund & 0 & 0 & 6 \\
&&&& \overline{\bf5}=\overline{\fund} & 0 & 0 & -6 \\
&&&& {\bf1} & 0 & 0 & 0 \\
& {\bf6}=\fund & 0 & 7 & {\bf5}=\fund & 0 & 7 & 1 \\
&&&& {\bf1} & 0 & 7 & -5 \\
& \overline{\bf6}=\overline{\fund} & 0 & -7 & \overline{\bf5}=\overline{\fund} & 0 & -7 & -1 \\
&&&& {\bf1} & 0 & -7 & 5 \\
& {\bf1} & 0 & 0 & {\bf1} & 0 & 0 & 0 \\\hline\end{array}
\ee
\caption{Decompositions of $U(7)\equiv SU(7)\times U(1)_E$ representations.\label{tab:SU7}}
\end{table}

\section{Fermion projectors\label{app:lambda}}

The projectors for the third generation of matter are explicitly given by
\begin{align}
(\lambda{}_\chi^{\bf10})_{i_4i_2}^{IJK} & =\frac{1}{\sqrt{6}}\lambda_\chi\delta{}^{IJK}_{i_4,i_2+3,7} &
(\lambda{}_\chi^{\bf5})_{i_5}^{IJK} & =\frac{1}{\sqrt{6}}\lambda_\chi\delta{}^{IJK}_{i_567} \nonumber\\
(\lambda_t)^{i_3i_2,IJK} & =\frac{1}{6\sqrt{6}}\lambda_t\epsilon^{i_3,i_2+3,k_5l_5m_5}\delta{}^{IJK}_{k_5l_5m_5} &
(\lambda_b)^{i_3i_2}_{IJK} & =\frac{1}{\sqrt{6}}\lambda_b\delta{}_{IJK}^{i_3,i_2+3,6} \nonumber\\
(\lambda_{t^c})_{i_3,IJK} & =\frac{1}{2\sqrt{6}}\epsilon_{i_3j_3k_3}\delta{}^{j_3k_37}_{IJK} & 
(\lambda_{b^c})_{i_3}^{IJK} & =\frac{1}{\sqrt{6}}\lambda_{b^c}\delta{}_{i_367}^{IJK} \nonumber\\
(\lambda_\nu)^{i_2,IJ} & =\frac{1}{\sqrt{2}}\lambda_\nu\epsilon^{i_2j_2}\delta{}^{IJ}_{j_26} &
(\lambda_\tau)^{i_2,IJ} & =\frac{1}{\sqrt{2}}\lambda_\tau\epsilon^{i_2j_2}\delta{}^{IJ}_{j_27} \nonumber\\
(\lambda_{N^c})_{IJ} & =\frac{1}{\sqrt{2}}\delta{}^{67}_{IJ} & 
(\lambda_{\tau^c})_{IJ} & =\frac{1}{\sqrt{2}}\lambda_{\tau^c}\delta{}^{45}_{IJ}
\end{align}
making use of the generalised Kronecker delta
\be
\delta{}_{ijk}^{lmn}=\norm{\begin{array}{rrr}\delta{}_i^l&\delta{}_i^m&\delta{}_i^n\\\delta{}_j^l&\delta{}_j^m&\delta{}_j^n\\\delta{}_k^l&\delta{}_k^m&\delta{}_k^n\end{array}}.
\ee
In all of these projectors $I=1,\ldots,7$ denote fundamental $SU(7)$ indices, $i_5=1,\ldots,5$ and $i_4=1,\ldots,4$ denote fundamental $SU(5)$ indices, and $i_3=1,2,3$ and $i_2=1,2$ denote fundamental SM $SU(3)$ and $SU(2)$ indices respectively.  The normalisations are chosen such that
\begin{align}
(\lambda{}_\chi^{{\bf10}*})^{i_4i_2}_{IJK}(\lambda{}_\chi^{\bf10})_{i_4i_2}^{IJK} & =7|\lambda_\chi|^2 &
(\lambda{}_\chi^{{\bf5}*})^{i_5}_{IJK}(\lambda{}_\chi^{\bf5})_{i_5}^{IJK} & =5|\lambda_\chi|^2 \nonumber\\
(\lambda_t^*)_{i_3i_2,IJK}(\lambda_t)^{i_3i_2,IJK} & =6|\lambda_t|^2 &
(\lambda_b^*)_{i_3i_2}^{IJK}(\lambda_b)^{i_3i_2}_{IJK} & =6|\lambda_b|^2 \nonumber\\
(\lambda_{t^c}^*)^{i_3,IJK}(\lambda_{t^c})_{i_3,IJK} & =3 &
(\lambda_{b^c}^*)^{i_3}_{IJK}(\lambda_{b^c})_{i_3}^{IJK} & =3|\lambda_{b^c}|^2 \nonumber\\
(\lambda_\nu^*)_{i_2,IJ}(\lambda_\nu)^{i_2,IJ} & =2|\lambda_\nu|^2 &
(\lambda_\tau^*)_{i_2,IJ}(\lambda_\tau)^{i_2,IJ} & =2|\lambda_\tau|^2 \nonumber\\
(\lambda_{N^c}^*)^{IJ}(\lambda_{N^c})_{IJ} & =|\lambda_{N^c}|^2 &
(\lambda_{\tau^c}^*)^{IJ}(\lambda_{\tau^c})_{IJ} & =|\lambda_{\tau^c}|^2
\end{align}
ensuring that the correct number of degrees of freedom propagate around the loops generating the pNGB potential.  $\lambda_{t^c}$ does not correspond to a mixing as the right-handed top quark is fully composite, hence it has no magnitude.  More generally one can choose different magnitudes for the individual components of each projector provided the SM gauge symmetry is respected.  We will stick with a single magnitude for each projector for simplicity.

\section{The pNGB potential\label{app:VpNGB}}

At quadratic order in the $\lambda$'s the pNGB potential is a function of the SM gauge singlet combinations
\begin{align}
[\lambda{}_\chi^2]_K^L & \equiv(\lambda{}_\chi^{{\bf10}*})^{i_4i_2}_{IJK}(\lambda{}_\chi^{\bf10})_{i_4i_2}^{IJL}+(\lambda{}_\chi^{{\bf5}*})^{i_5}_{IJK}(\lambda{}_\chi^{\bf5})_{i_5}^{IJL} &
[\lambda{}_t^2]_K^L & \equiv(\lambda_t^*)_{i_3i_2,IJK}(\lambda_t)^{i_3i_2,IJL} \nonumber\\
[\lambda{}_b^2]_K^L & \equiv(\lambda_b^*)_{i_3i_2}^{IJL}(\lambda_b)^{i_3i_2}_{IJK} &
[\lambda_{b^c}{}^2]_K^L & \equiv(\lambda_{b^c}{}^*)^{i_3}_{IJK}(\lambda_{b^c})_{i_3}^{IJL}.
\end{align}
At one loop in the elementary matter fields there are then four contributions to the pNGB potential
\be
V{}_1^\psi=F{}^\psi_1[\lambda{}_\psi^2]_K^L\Sigma{}^K_L
\ee
for $\psi=\chi,t,b,b^c$, these being generated by diagrams like that shown in figure~\ref{fig:lambda2}.

The $F$'s come from the strong sector form factors
\be
F{}_1^\psi=2\int\frac{\mathrm{d}^{4}p}{(2\pi)^4}\,\Pi^\psi(p)=2c{}_1^\psi\frac{1}{16\pi^2}\frac{m_\rho^4}{g_\rho^2}=2c{}_1^\psi\frac{g_\rho^2}{16\pi^2}f^4
\ee
using $m_\rho=g_\rho f$ and where the $c{}_1$'s  are unknown, order-one coefficients.  To arrive at the above expression we have cutoff the momentum integration at $m_\rho$ and multiplied by a loop factor of $1/(16\pi^2)$.  There is a further suppression by a factor of $1/g_\rho^2$ such that $V\sim m_\rho^4/(16\pi^2)$ in the strong coupling limit $\lambda\sim g_\rho$.  The overall factor of two comes from the two fermion polarisations propagating around the loop.

Other contributions to the pNGB potential are suppressed by powers of $|\lambda|/g_\rho$, which must be small for the global symmetry of the strong sector to remain approximately preserved, or by elementary sector loop factors.  Adding up all contributions, using eqs.~\eqref{eq:Sigma7} and \eqref{eq:Sigmau} to substitute in the components of $\Sigma$, then splitting $H$ into its doublet and triplet components, $D$ and $T$ respectively, we find
\begin{align}
V_{\rm matter}={} & \frac{1}{3f^2}F{}_1^\chi|\lambda_\chi|^2\nbrack{12-9|T|^2-7|D|^2-7|S|^2}+\frac{1}{3f^2}F{}_1^t|\lambda_t|^2\nbrack{4|T|^2+3|D|^2}{} \nonumber\\
&+\frac{1}{3f^2}F{}_1^b|\lambda_b|^2\nbrack{2|T|^2+3|D|^2+6|S|^2}+\frac{1}{3f^2}F{}_1^{b^c}|\lambda_{b^c}|^2\nbrack{3-2|T|^2-3|D|^2}.
\end{align}

The elementary gauge field contribution to the pNGB potential stems from the terms in the effective Lagrangian given in eq.~\eqref{eq:Leffg}.  At one loop in the elementary gauge fields and at leading (quadratic) order in the $\Omega$'s there are two contributions to the pNGB potential
\begin{align}
V{}_1^{\rm A} & =F{}^{\rm A}_1\sbrack{\Omega{}_A^a(T^A)_I^J(T^B)_K^I\Omega{}_B^a}\Sigma_J^K \nonumber\\
V{}_2^{\rm A} & =F{}^{\rm A}_2\sbrack{\Omega{}_A^a(T^A)_I^J(T^B)_K^L\Omega{}_B^a}\Sigma_J^K\Sigma_L^I
\end{align}
these being generated by diagrams like that shown in figure~\ref{fig:omega2}.  The $F$'s are derived from the strong sector form factors as before
\be
F{}_{1,2}^{\rm A}=3\int\frac{\mathrm{d}^{4}p}{(2\pi)^4}\,\frac{\Pi{}_{1,2}^{\rm A}(p)}{p^2}=3c{}_{1,2}^{\rm A}\frac{1}{16\pi^2}\frac{m_\rho^4}{g_\rho^2}=3c{}_{1,2}^{\rm A}\frac{g_\rho^2}{16\pi^2}f^4
\ee
but now with a factor of three for the three gauge field polarisations propagating around the loop.  Using the definition of the projector in eq.~\eqref{eq:Omegadef} these can be written in terms of $SU(3)\times SU(2)$ components
\begin{align}
V{}_1^{\rm A} & =\frac{1}{f^2}F{}^{\rm A}_1\sbrack{g_3^2(T^{a_3})_{i_3}^{j_3}(T^{a_3})_{k_3}^{i_3}T_{j_3}T^{\dag k_3}+g_2^2(T^{a_2})_{i_2}^{j_2}(T^{a_2})_{k_2}^{i_2}D_{j_2}D^{\dag k_2}} \nonumber\\
V{}_2^{\rm A} & =\frac{1}{f^4}F{}^{\rm A}_2\sbrack{g_3^2(T^{a_3})_{i_3}^{j_3}(T^{a_3})_{k_3}^{l_3}T_{j_3}T_{l_3}T^{\dag i_3}T^{\dag k_3}+g_2^2(T^{a_2})_{i_2}^{j_2}(T^{a_2})_{k_2}^{l_2}D_{j_2}D_{l_2}D^{\dag i_2}D^{\dag k_2}}
\end{align}
neglecting the much smaller hypercharge contributions (hypercharge only contributes terms that are functions of $|T|^2$, $|D|^2$ and $|S|^2$ so has no effect on the symmetries respected by the potential).  The overall contribution from elementary gauge fields is therefore
\be
V_{\rm gauge}=\frac{1}{f^2}F{}^{\rm A}_1\nbrack{\frac{4}{3}g_3^2|T|^2+\frac{3}{4}g_2^2|D|^2}+\frac{1}{f^4}F{}^{\rm A}_2\nbrack{\frac{1}{3}g_3^2|T|^4+\frac{1}{4}g_2^2|D|^4}.
\ee

At quartic order in the $\lambda$'s there are also important contributions to the pNGB potential.  These are functions of the SM gauge singlet combinations of the projectors
\begin{align}
[\lambda_\chi^2]_{IJK}^{MNP} & \equiv(\lambda{}_\chi^{{\bf10}*})^{i_4i_2}_{IJK}(\lambda{}_\chi^{\bf10})_{i_4i_2}^{MNP}+(\lambda{}_\chi^{{\bf5}*})^{i_5}_{IJK}(\lambda{}_\chi^{\bf5})_{i_5}^{MNP} &
[\lambda_t^2]^{MNP}_{IJK} & \equiv(\lambda_t^*)_{i_3i_2,IJK}(\lambda_t)^{i_3i_2,MNP} \nonumber\\
[\lambda_b^2]_{IJK}^{MNP} & \equiv(\lambda_b^*)_{i_3i_2}^{MNP}(\lambda_b)^{i_3i_2}_{IJK} &
[\lambda_{b^c}{}^2]^{MNP}_{IJK} & \equiv(\lambda_{b^c}{}^*)^{i_3}_{IJK}(\lambda_{b^c})_{i_3}^{MNP}
\end{align}
and
\be
[\lambda_{tb}]^{IJKMNP}\equiv(\lambda_t)^{i_3i_2,IJK}(\lambda_b^*)_{i_3i_2}^{MNP}.
\ee
At one loop in the elementary matter fields there are seven contributions to the pNGB potential
\be
V{}_2^{\psi\psi^\prime}=\left\{\begin{array}{ll}
F{}^{\psi\psi^\prime}_2[\lambda{}_\psi^2]_{IJK}^{MNP}\Sigma{}_P^O[\lambda{}_{\psi^\prime}^2]^{IJL}_{MNO}\Sigma{}_L^K & \mbox{for }\psi\psi^\prime=\chi\chi,tt,bb,b^cb^c \\
2F{}^{\psi\psi^\prime}_2[\lambda{}_\chi^2]_{IJK}^{MNP}\Sigma{}_P^O[\lambda{}_q^2]^{IJL}_{MNO}\Sigma{}_L^K & \mbox{for }\psi\psi^\prime=\chi t,bb^c \\
F{}^{tb}_2[\lambda_{tb}]^{IJLMNP}[\lambda_{tb}^*]_{IJKMNO}\Sigma{}_L^K\Sigma{}_P^O
\end{array}\right.
\ee
which are generated by diagrams like the one in figure~\ref{fig:lambda4}.  As before, the $F$'s come from strong sector form factors
\be
F{}_2^{\psi\psi^\prime}=2\int\frac{\mathrm{d}^{4}p}{(2\pi)^4}\,\sbrack{\Pi^{\psi\psi^\prime}(p)}^2=2c{}_2^{\psi\psi^\prime}\frac{1}{16\pi^2}\frac{m_\rho^4}{g_\rho^4}=2c{}_2^{\psi\psi^\prime}\frac{1}{16\pi^2}f^4
\ee
for $\psi\psi^\prime=(\chi\chi,tt,bb,b^cb^c,\chi t)$
\be
F{}_2^{bb^c}=2\int\frac{\mathrm{d}^{4}p}{(2\pi)^4}\,\frac{1}{p^2}\sbrack{M^{bb^c}(p)}^2=2c{}_2^{bb^c}\frac{1}{16\pi^2}\frac{m_\rho^4}{g_\rho^4}=2c{}_2^{bb^c}\frac{1}{16\pi^2}f^4
\ee
and
\be
F{}_2^{tb}=2\int\frac{\mathrm{d}^{4}p}{(2\pi)^4}\,\Pi^t(p)\Pi^b(p)=2c{}_2^{tb}\frac{1}{16\pi^2}\frac{m_\rho^4}{g_\rho^4}=2c{}_2^{tb}\frac{1}{16\pi^2}f^4.
\ee
Collecting all relevant terms we find
\begin{align}
V\supset{} & {}\frac{1}{16\pi^2}\nbrack{\frac{28}{9}c{}_2^{\chi\chi}|\lambda_\chi|^4+\frac{4}{3}c{}_2^{bb}|\lambda_b|^4-\frac{4}{3}c{}_2^{bb^c}|\lambda_b|^2|\lambda_{b^c}|^2+\frac{2}{3}c{}_2^{tb}|\lambda_t|^2|\lambda_b|^2}|D|^2|S|^2
\end{align}
up to quadratic order in the scalar fields and to leading order in $v/f$.

\bibliographystyle{JHEP-2}
\bibliography{SRS}

\providecommand{\href}[2]{#2}\begingroup\raggedright\begin{thebibliography}{10}

\bibitem{Kaplan:1983fs}
D.~B. Kaplan and H.~Georgi, {\it {$SU(2)\times U(1)$ Breaking by Vacuum
  Misalignment}},  {\em Phys.Lett.} {\bf B136} (1984) 183.

\bibitem{Kaplan:1983sm}
D.~B. Kaplan, H.~Georgi, and S.~Dimopoulos, {\it {Composite Higgs Scalars}},
  {\em Phys.Lett.} {\bf B136} (1984) 187.

\bibitem{Dugan:1984hq}
M.~J. Dugan, H.~Georgi, and D.~B. Kaplan, {\it {Anatomy of a Composite Higgs
  Model}},  {\em Nucl.Phys.} {\bf B254} (1985) 299.

\bibitem{Contino:2003ve}
R.~Contino, Y.~Nomura, and A.~Pomarol, {\it {Higgs as a holographic pseudo
  Goldstone boson}},  {\em Nucl.Phys.} {\bf B671} (2003) 148--174,
  [\href{http://xxx.lanl.gov/abs/hep-ph/0306259}{{\tt hep-ph/0306259}}].

\bibitem{Agashe:2004rs}
K.~Agashe, R.~Contino, and A.~Pomarol, {\it {The Minimal composite Higgs
  model}},  {\em Nucl.Phys.} {\bf B719} (2005) 165--187,
  [\href{http://xxx.lanl.gov/abs/hep-ph/0412089}{{\tt hep-ph/0412089}}].

\bibitem{Agashe:2003zs}
K.~Agashe, A.~Delgado, M.~J. May, and R.~Sundrum, {\it {RS1, custodial isospin
  and precision tests}},  {\em JHEP} {\bf 0308} (2003) 050,
  [\href{http://xxx.lanl.gov/abs/hep-ph/0308036}{{\tt hep-ph/0308036}}].

\bibitem{Rattazzi:2000hs}
R.~Rattazzi and A.~Zaffaroni, {\it {Comments on the holographic picture of the
  Randall-Sundrum model}},  {\em JHEP} {\bf 0104} (2001) 021,
  [\href{http://xxx.lanl.gov/abs/hep-th/0012248}{{\tt hep-th/0012248}}].

\bibitem{Cacciapaglia:2007fw}
G.~Cacciapaglia, C.~Csaki, J.~Galloway, G.~Marandella, J.~Terning, {\em
  et.~al.}, {\it {A GIM Mechanism from Extra Dimensions}},  {\em JHEP} {\bf
  0804} (2008) 006, [\href{http://xxx.lanl.gov/abs/0709.1714}{{\tt
  arXiv:0709.1714}}].

\bibitem{Santiago:2008vq}
J.~Santiago, {\it {Minimal Flavor Protection: A New Flavor Paradigm in Warped
  Models}},  {\em JHEP} {\bf 0812} (2008) 046,
  [\href{http://xxx.lanl.gov/abs/0806.1230}{{\tt arXiv:0806.1230}}].

\bibitem{Wells:2003tf}
J.~D. Wells, {\it {Implications of supersymmetry breaking with a little
  hierarchy between gauginos and scalars}},
  \href{http://xxx.lanl.gov/abs/hep-ph/0306127}{{\tt hep-ph/0306127}}.

\bibitem{ArkaniHamed:2004fb}
N.~Arkani-Hamed and S.~Dimopoulos, {\it {Supersymmetric unification without low
  energy supersymmetry and signatures for fine-tuning at the LHC}},  {\em JHEP}
  {\bf 0506} (2005) 073, [\href{http://xxx.lanl.gov/abs/hep-th/0405159}{{\tt
  hep-th/0405159}}].

\bibitem{Arvanitaki:2012ps}
A.~Arvanitaki, N.~Craig, S.~Dimopoulos, and G.~Villadoro, {\it {Mini-Split}},
  {\em JHEP} {\bf 1302} (2013) 126,
  [\href{http://xxx.lanl.gov/abs/1210.0555}{{\tt arXiv:1210.0555}}].

\bibitem{ArkaniHamed:2012gw}
N.~Arkani-Hamed, A.~Gupta, D.~E. Kaplan, N.~Weiner, and T.~Zorawski, {\it
  {Simply Unnatural Supersymmetry}},
  \href{http://xxx.lanl.gov/abs/1212.6971}{{\tt arXiv:1212.6971}}.

\bibitem{Kaplan:1991dc}
D.~B. Kaplan, {\it {Flavor at SSC energies: A New mechanism for dynamically
  generated fermion masses}},  {\em Nucl.Phys.} {\bf B365} (1991) 259--278.

\bibitem{Gherghetta:2000qt}
T.~Gherghetta and A.~Pomarol, {\it {Bulk fields and supersymmetry in a slice of
  AdS}},  {\em Nucl.Phys.} {\bf B586} (2000) 141--162,
  [\href{http://xxx.lanl.gov/abs/hep-ph/0003129}{{\tt hep-ph/0003129}}].

\bibitem{Agashe:2005vg}
K.~Agashe, R.~Contino, and R.~Sundrum, {\it {Top compositeness and precision
  unification}},  {\em Phys.Rev.Lett.} {\bf 95} (2005) 171804,
  [\href{http://xxx.lanl.gov/abs/hep-ph/0502222}{{\tt hep-ph/0502222}}].

\bibitem{Agashe:2004ci}
K.~Agashe and G.~Servant, {\it {Warped unification, proton stability and dark
  matter}},  {\em Phys.Rev.Lett.} {\bf 93} (2004) 231805,
  [\href{http://xxx.lanl.gov/abs/hep-ph/0403143}{{\tt hep-ph/0403143}}].

\bibitem{Agashe:2004bm}
K.~Agashe and G.~Servant, {\it {Baryon number in warped GUTs: Model building
  and (dark matter related) phenomenology}},  {\em JCAP} {\bf 0502} (2005) 002,
  [\href{http://xxx.lanl.gov/abs/hep-ph/0411254}{{\tt hep-ph/0411254}}].

\bibitem{Frigerio:2012uc}
M.~Frigerio, A.~Pomarol, F.~Riva, and A.~Urbano, {\it {Composite Scalar Dark
  Matter}},  {\em JHEP} {\bf 1207} (2012) 015,
  [\href{http://xxx.lanl.gov/abs/1204.2808}{{\tt arXiv:1204.2808}}].

\bibitem{Frigerio:2011zg}
M.~Frigerio, J.~Serra, and A.~Varagnolo, {\it {Composite GUTs: models and
  expectations at the LHC}},  {\em JHEP} {\bf 1106} (2011) 029,
  [\href{http://xxx.lanl.gov/abs/1103.2997}{{\tt arXiv:1103.2997}}].

\bibitem{Vecchi:2013iza}
L.~Vecchi, {\it {WIMPs and Un-Naturalness}},
  \href{http://xxx.lanl.gov/abs/1312.5695}{{\tt arXiv:1312.5695}}.

\bibitem{Mrazek:2011iu}
J.~Mrazek, A.~Pomarol, R.~Rattazzi, M.~Redi, J.~Serra, {\em et.~al.}, {\it {The
  Other Natural Two Higgs Doublet Model}},  {\em Nucl.Phys.} {\bf B853} (2011)
  1--48, [\href{http://xxx.lanl.gov/abs/1105.5403}{{\tt arXiv:1105.5403}}].

\bibitem{Contino:2010rs}
R.~Contino, {\it {The Higgs as a Composite Nambu-Goldstone Boson}},
  \href{http://xxx.lanl.gov/abs/1005.4269}{{\tt arXiv:1005.4269}}.

\bibitem{Bertuzzo:2012ya}
E.~Bertuzzo, T.~S. Ray, H.~de~Sandes, and C.~A. Savoy, {\it {On Composite Two
  Higgs Doublet Models}},  {\em JHEP} {\bf 1305} (2013) 153,
  [\href{http://xxx.lanl.gov/abs/1206.2623}{{\tt arXiv:1206.2623}}].

\bibitem{Bellazzini:2014yua}
B.~Bellazzini, C.~Csaki, and J.~Serra, {\it {Composite Higgses}},  {\em
  Eur.Phys.J.} {\bf C74} (2014) 2766,
  [\href{http://xxx.lanl.gov/abs/1401.2457}{{\tt arXiv:1401.2457}}].

\bibitem{Agashe:2006iy}
K.~Agashe, A.~E. Blechman, and F.~Petriello, {\it {Probing the Randall-Sundrum
  geometric origin of flavor with lepton flavor violation}},  {\em Phys.Rev.}
  {\bf D74} (2006) 053011, [\href{http://xxx.lanl.gov/abs/hep-ph/0606021}{{\tt
  hep-ph/0606021}}].

\bibitem{KerenZur:2012fr}
B.~Keren-Zur, P.~Lodone, M.~Nardecchia, D.~Pappadopulo, R.~Rattazzi, {\em
  et.~al.}, {\it {On Partial Compositeness and the CP asymmetry in charm
  decays}},  {\em Nucl.Phys.} {\bf B867} (2013) 394--428,
  [\href{http://xxx.lanl.gov/abs/1205.5803}{{\tt arXiv:1205.5803}}].

\bibitem{Choi:2002ps}
K.-w. Choi and I.-W. Kim, {\it {One loop gauge couplings in $AdS_5$}},  {\em
  Phys.Rev.} {\bf D67} (2003) 045005,
  [\href{http://xxx.lanl.gov/abs/hep-th/0208071}{{\tt hep-th/0208071}}].

\bibitem{Gherghetta:2004sq}
T.~Gherghetta, {\it {Partly supersymmetric grand unification}},  {\em
  Phys.Rev.} {\bf D71} (2005) 065001,
  [\href{http://xxx.lanl.gov/abs/hep-ph/0411090}{{\tt hep-ph/0411090}}].

\bibitem{Barnard:2013zea}
J.~Barnard, T.~Gherghetta, and T.~S. Ray, {\it {UV descriptions of composite
  Higgs models without elementary scalars}},  {\em JHEP} {\bf 1402} (2014) 002,
  [\href{http://xxx.lanl.gov/abs/1311.6562}{{\tt arXiv:1311.6562}}].

\bibitem{Marzocca:2012zn}
D.~Marzocca, M.~Serone, and J.~Shu, {\it {General Composite Higgs Models}},
  {\em JHEP} {\bf 1208} (2012) 013,
  [\href{http://xxx.lanl.gov/abs/1205.0770}{{\tt arXiv:1205.0770}}].

\bibitem{Pomarol:2012qf}
A.~Pomarol and F.~Riva, {\it {The Composite Higgs and Light Resonance
  Connection}},  {\em JHEP} {\bf 1208} (2012) 135,
  [\href{http://xxx.lanl.gov/abs/1205.6434}{{\tt arXiv:1205.6434}}].

\bibitem{Cline:2013gha}
J.~M. Cline, K.~Kainulainen, P.~Scott, and C.~Weniger, {\it {Update on scalar
  singlet dark matter}},  {\em Phys.Rev.} {\bf D88} (2013) 055025,
  [\href{http://xxx.lanl.gov/abs/1306.4710}{{\tt arXiv:1306.4710}}].

\bibitem{Aprile:2012nq}
{\bf XENON100} Collaboration, {\it {Dark Matter Results from 225 Live Days of
  XENON100 Data}},  {\em Phys.Rev.Lett.} {\bf 109} (2012) 181301,
  [\href{http://xxx.lanl.gov/abs/1207.5988}{{\tt arXiv:1207.5988}}].

\bibitem{Aad:2012pra}
{\bf ATLAS} Collaboration, {\it {Searches for heavy long-lived sleptons and
  R-Hadrons with the ATLAS detector in $pp$ collisions at $\sqrt{s}=7$ TeV}},
  {\em Phys.Lett.} {\bf B720} (2013) 277--308,
  [\href{http://xxx.lanl.gov/abs/1211.1597}{{\tt arXiv:1211.1597}}].

\bibitem{Chatrchyan:2013oca}
{\bf CMS} Collaboration, {\it {Searches for long-lived charged particles in pp
  collisions at $\sqrt{s}$=7 and 8 TeV}},  {\em JHEP} {\bf 1307} (2013) 122,
  [\href{http://xxx.lanl.gov/abs/1305.0491}{{\tt arXiv:1305.0491}}].

\bibitem{Aad:2014lra}
{\bf ATLAS} Collaboration, {\it {Search for strong production of supersymmetric
  particles in final states with missing transverse momentum and at least three
  b-jets at $\sqrt{s} =$ 8 TeV proton-proton collisions with the ATLAS
  detector}},  \href{http://xxx.lanl.gov/abs/1407.0600}{{\tt arXiv:1407.0600}}.

\bibitem{2012arXiv1206.6379F}
R.~Feger and T.~W. Kephart, {\it {LieART -- A Mathematica application for Lie
  algebras and representation theory}},
  \href{http://xxx.lanl.gov/abs/1206.6379}{{\tt arXiv:1206.6379}}.

\end{thebibliography}\endgroup
\end{document}